\documentclass[amsmath,amssymb,superscriptaddress,aps, prb,twocolumn]{revtex4-1}

\usepackage[T1]{fontenc}
\usepackage[utf8]{inputenc}
\usepackage{xcolor}
\usepackage{siunitx}
\usepackage{graphicx}
\usepackage{hyperref}

\begin{document}
\title{Superconducting fluctuations in a thin NbN film probed by the Hall effect}

\author{Daniel Destraz}
\email{destraz@physik.uzh.ch}
\affiliation{Physik-Institut, Universität Zürich, Winterthurerstrasse 190, CH-8057 Zürich, Switzerland}

\author{Konstantin Ilin}
\affiliation{Institut für Mikro- und Nanoelektronische Systeme (IMS), Karlsruher Institut für Technologie, Germany}

\author{Michael Siegel}
\affiliation{Institut für Mikro- und Nanoelektronische Systeme (IMS), Karlsruher Institut für Technologie, Germany}

\author{Andreas Schilling}
\affiliation{Physik-Institut, Universität Zürich, Winterthurerstrasse 190, CH-8057 Zürich, Switzerland}

\author{Johan Chang}
\affiliation{Physik-Institut, Universität Zürich, Winterthurerstrasse 190, CH-8057 Zürich, Switzerland}

\begin{abstract}
We present a comprehensive study of how  superconducting fluctuations  in the normal state contribute to the conductivity tensor in a thin (\SI{119}{\angstrom}) film of NbN. It is shown how these fluctuations drive a sign change in the Hall coefficient $R_\mathrm{H}$ for low magnetic fields near the superconducting transition. The scaling behaviours as a function of distance to the  transition $\epsilon=\ln(T/T_\mathrm{c})$ of the longitudinal ($\sigma_\mathrm{xx}$) and transverse ($\sigma_\mathrm{xy}$) conductivity is found to be consistent with Gaussian fluctuation theory. Moreover, excellent quantitative agreement between theory and experiment is obtained without any adjustable parameters. Our experimental results thus provide a case study of the conductivity tensor originating from short-lived Cooper pairs.
\end{abstract}
\maketitle

\section{Introduction}
Superconducting fluctuations in the normal state have long been known to influence bulk properties such as conductivity and magnetization. Generally, a stronger response to fluctuations is expected for lower dimensions \cite{UssishkinPRL02}. Many studies have therefore been carried out on thin films or layered compounds such as the high-temperature cuprate superconductors~\cite{EmeryNature1995}. Of particular interest are systems that host a Bardeen--Cooper--Schrieffer (BCS) regime to Bose--Einstein condensate (BEC) cross-overs~\cite{RanderiaNatPhys2010,GantmakherPU10}. The BCS regime is characterized by conventional Gaussian fluctuations~\cite{UssishkinPRL02,MichaeliEPL09,SerbynPRL09} whereas  phase fluctuations are expected in the BEC regime. In the cuprates paraconductivity, torque magnetization, and Nernst effect experiments -- inside the pseudogap phase -- have been interpreted as evidence for phase fluctuations of superconductivity~\cite{XuNat00,WangPRB03,LiPRB2010,BaityPRB2016}. The topic, however, remains controversial as the same techniques have also produced results consistent with Gaussian fluctuation theory~\cite{ChangNatPhys2012,PourretNatPhys2006,CaboPRL07,RullierAlbenquePRB2011,LeridonPRB2007,PerfettiPRL15,BehniaRPP16}. To make progress, one way forward is to study superconducting fluctuations of related systems. Recently, a pseudogap phase has been identified in  disordered films of NbN and TiN and it has been conjectured that it stems from phase fluctuating superconductivity~\cite{SacepeNATCOMM10,MondalPRL11,ChandPRB2012}. In this context, careful studies of the normal state fluctuations are called for. Recently, the sister compound TaN, for which no pseudogap has been identified, has been studied and it was demonstrated that superconducting fluctuations manifest themselves in the Hall coefficient~\cite{BreznayPRB2012,BreznayPRB13} -- consistent with predictions of Gaussian fluctuation theory~\cite{MichaeliPRB2012}.

Due to its promising potential for applications such as single-photon detection and hot-electron bolometers, NbN is one of the best characterized superconducting films. Both the normal state metallic and the superconducting properties have been widely studied~\cite{ChockalingamPRB2008,ChandPRB09,ChandPRB2012,SemenovPRB2009}. Perhaps for this reason, it has served as a model system for studies of out-of-equilibrium dynamics of  superconductivity~\cite{BeckPRL11,MatsunagaPRL12,BeckPRL13,MatsunagaSCIENCE14}. Here, we use a NbN film just outside the pseudogap regime to carry out  a combined paraconductivity and Hall effect study of the normal state superconducting fluctuations. The sign of the contribution from superconducting fluctuations to the Hall conductivity is defined by the derivative $\kappa = -\mathrm{d}\ln(T_\mathrm{c})/\mathrm{d} \mu$ where $\mu$ is the chemical potential~\cite{AronovPRB95,MichaeliPRB2012}. For most conventional superconductors, including NbN, $\kappa<0$ and hence the Hall coefficient due to Gaussian fluctuations is expected to be positive [$R_\mathrm{H}(\mathrm{SC})>0$]. It is also known that charge transport in NbN films is governed by electron-like carriers~\cite{ChockalingamPRB2008}, which generate a negative normal state quasiparticle (QP) Hall coefficient [$R_\mathrm{H}(\mathrm{QP})<0$]. In NbN short lived Cooper pairs and quasiparticles are thus expected to contribute with opposite sign to the Hall effect. Near $T_\mathrm{c}$, but still within the normal state ($T>T_\mathrm{c}$), we indeed find a sign reversal of the Hall effect response. This sign reversal facilitates the disentanglement of the Hall signal from quasiparticles and short lived Cooper pairs and hence enables us to study the response from superconducting (SC) fluctuations to the Hall effect. Although the Hall conductivity $\Delta \sigma_\mathrm{xy}$ generated by SC fluctuations is generally highly non-linear, it does scale with magnetic field $B$ in the limit $B\rightarrow 0$. Consistent with Gaussian fluctuation theory, $\Delta \sigma_\mathrm{xy} / B \propto \epsilon^{-2}$ scales with $\epsilon= \ln(T/T_\mathrm{c})$ that for $\epsilon\ll 1$ is a measure of the distance to the superconducting transition $(T-T_\mathrm{c}) / T_\mathrm{c}$. Furthermore, from the normal state Hall conductivity isotherms we extract a ghost critical field $B^*\propto\epsilon$. This combined with a paraconductivity  that scales as $\Delta \sigma_\mathrm{xx} \propto \epsilon^{-1}$ makes a convincing case for Gaussian fluctuations in NbN. Furthermore, excellent quantitative agreement between Gaussian fluctuation theory and the experiment is found without any adjustable parameters. Our study therefore provides an experimental demonstration of how Gaussian fluctuations of superconductivity contribute to the conductivity tensor.

\section{Methods} 
A thin film of NbN ($T_\mathrm{c} =$ \SI{14.96}{\K}) was grown on a sapphire substrate using dc reactive magnetron sputtering of a pure Nb target in a mixture of Ar and N$_2$ gasses. The average thickness $d=$ \SI{119(2)}{\angstrom} was measured with a stylus profiler. Six gold contacts with Hall bar geometry were deposited onto the film. Resistivity and Hall effect experiments were carried out -- using a commercial Quantum Design PPMS -- in magnetic fields up to 9 Tesla. Magnetic field and temperature were stabilized before measuring. Reversal of the field direction was used to eliminate contributions originating from contact misalignment.

\section{Results}
\begin{figure}
\includegraphics[width=\linewidth]{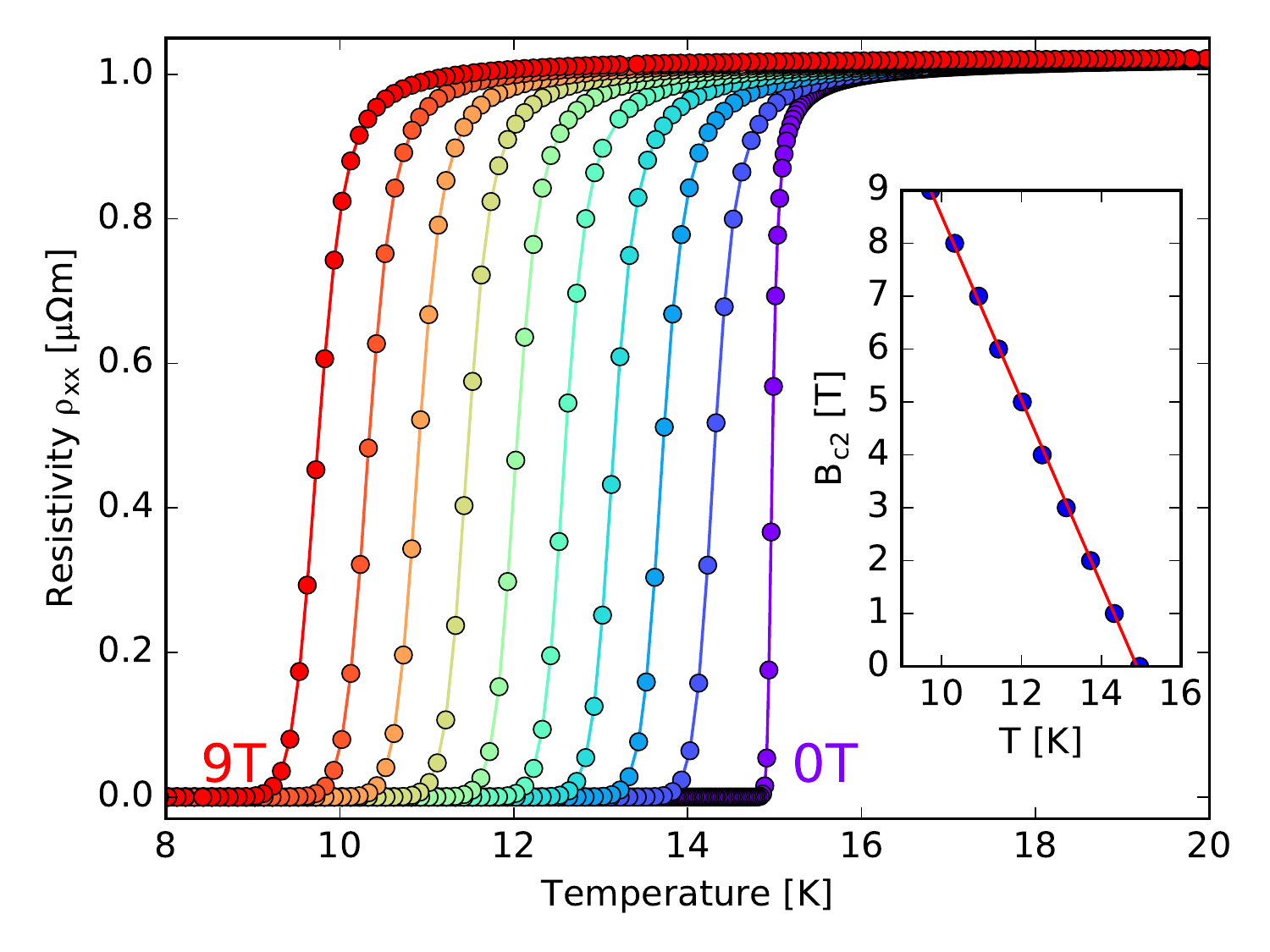}
\caption{\label{fig:R(T)} In-plane resistivity of a \SI{119}{\angstrom} thin NbN film as a function of temperature for magnetic fields in steps of \SI{1}{T}. The magnetic field is applied perpendicular to the film plane. Solid lines are guides to the eye. The upper critical field $B_\mathrm{c2}(T)$ (shown in the inset) is defined by the point with the steepest slope on the respective transitions. The red line in the inset is a linear fit used to evaluate $B_\mathrm{c2}(0)$ -- see main text for further explanation.}
\end{figure}

In Fig.~\ref{fig:R(T)} the longitudinal resistivity $\rho_\mathrm{xx}$ is shown as a function of temperature and magnetic field perpendicular to the film. The zero-field resistivity curve yields $T_\mathrm{c}=$ \SI{14.96}{\kelvin}, defined by the temperature with the largest derivative $\mathrm{d}\rho_\mathrm{xx}/\mathrm{d}T$. Notice that the sharpness of the transition allows determination of $T_\mathrm{c}$ with \SI{20}{\milli \kelvin} precision. When a magnetic field $B$ is applied perpendicular to the film, the transition temperature is gradually suppressed as indicated in the inset of Fig.~\ref{fig:R(T)}.

\begin{figure}
\includegraphics[width=\linewidth]{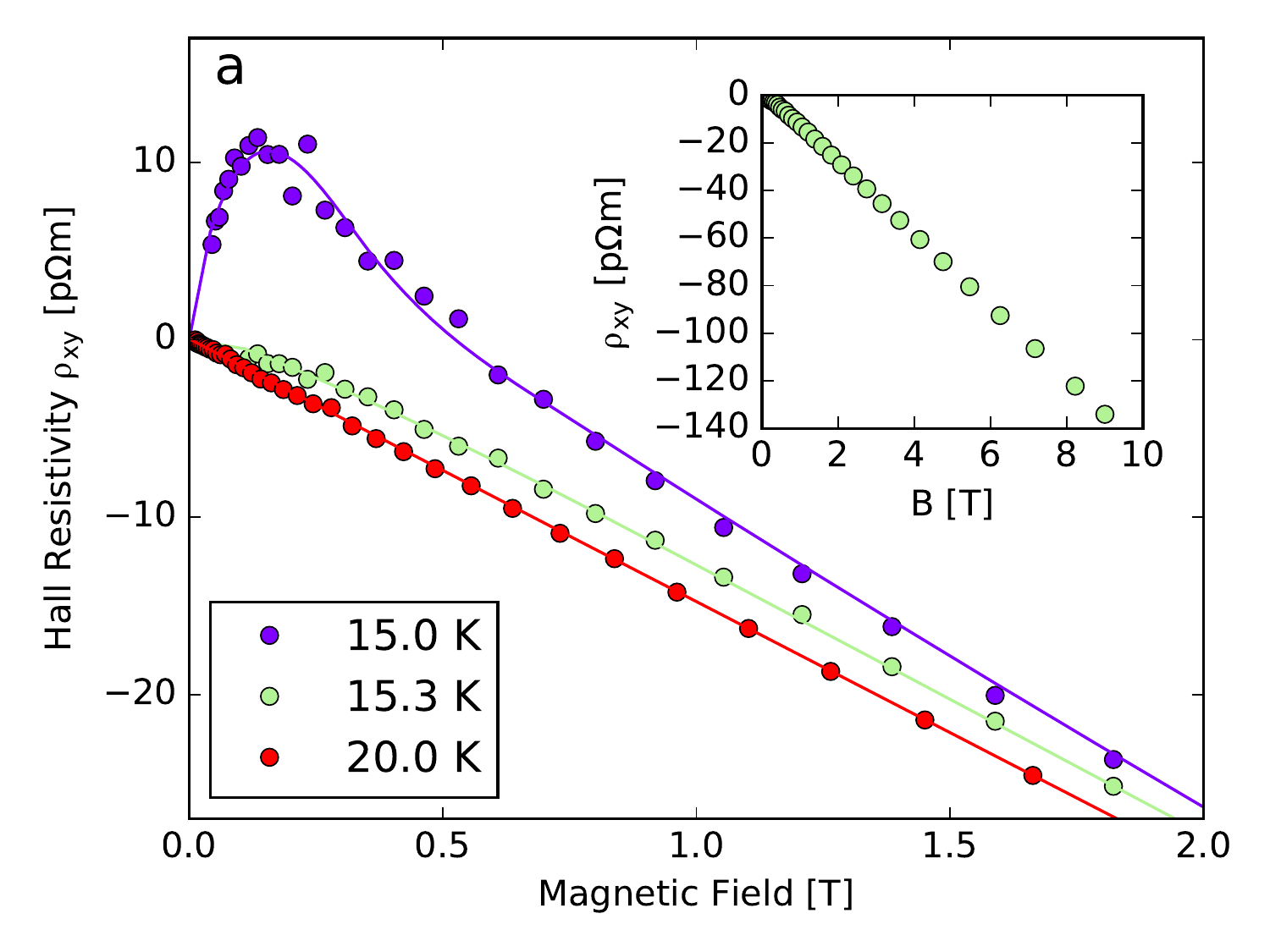}
\includegraphics[width=\linewidth]{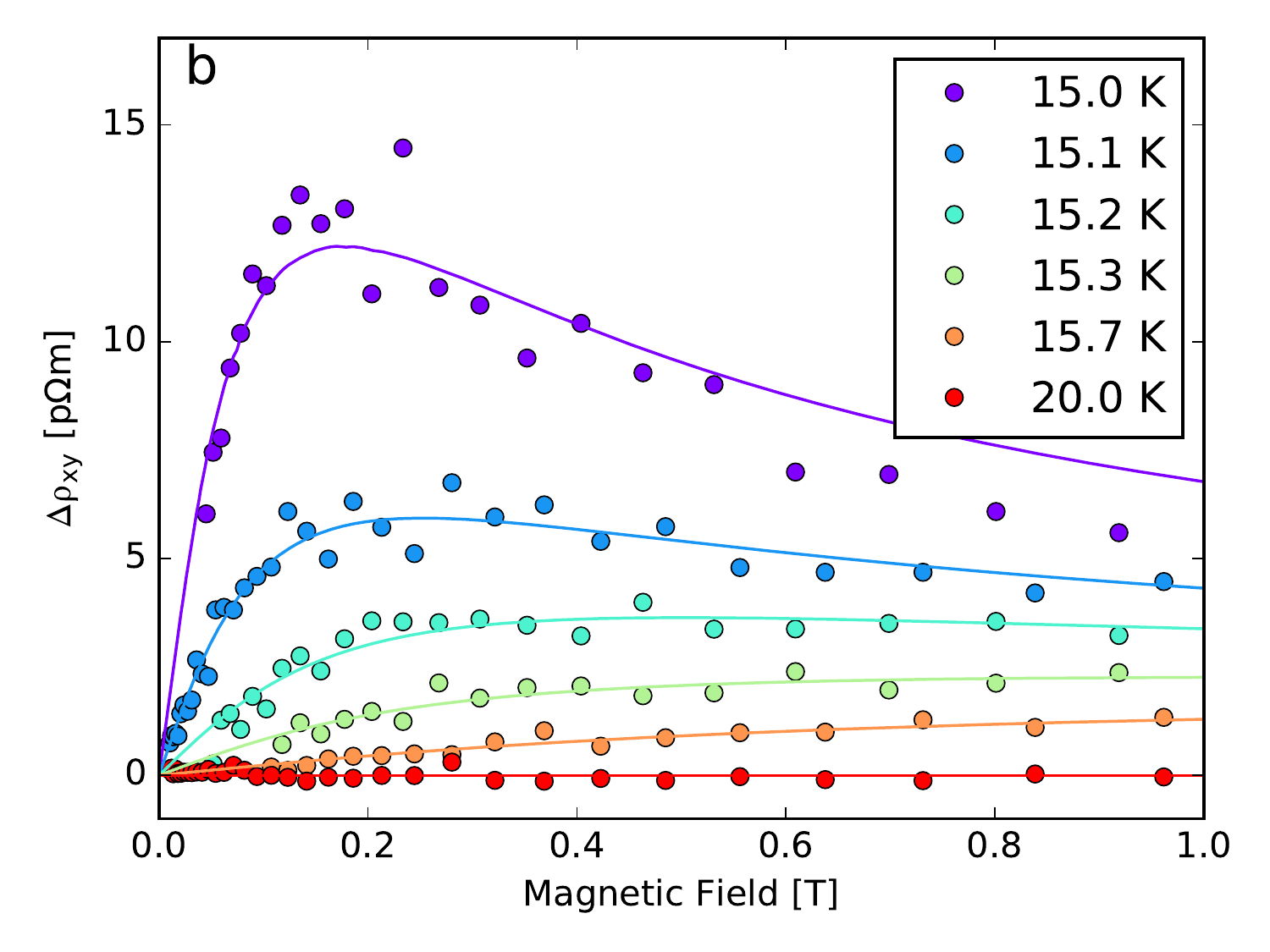}
\caption{\label{fig:R_xy}(a) Hall resistivity $\rho_\mathrm{xy}$  isotherms for $T \approx T_\mathrm{c}$, 15.3  and  \SI{20.0}{\K}. The inset displays the high-field linear field dependence of $\rho_\mathrm{xy}$.
(b) The non-linear Hall resistance $\Delta \rho_\mathrm{xy}$ obtained by subtracting the linear high-field dependence for each of the respective isotherms. Solid lines are guides to the eye.}
\end{figure}

Raw Hall resistivity ($\rho_\mathrm{xy}$) isotherms are shown in Fig.~\ref{fig:R_xy}a. Well above the superconducting transition, the negative Hall response $\rho_\mathrm{xy}$ scales linearly with  magnetic field strength. Essentially no magnetoresistance is observed in   $\rho_\mathrm{xx}$ and $R_\mathrm{H}=\rho_\mathrm{xy}/B$. This is consistent with a single band picture where the Hall coefficient is given by $R_\mathrm{H}= - 1 / (n e)$. Our film has a carrier density $n=$ \SI{4.2e23}{\cm^{-3}} and hence $k_\mathrm{F} = (3 \pi^2 n)^{1/3} $= 2.3~\AA$^{-1}$.  The electronic mean free path  $\ell = \hbar k_\mathrm{F} / ne^2 \rho =$ 2.0~\AA\ confirms that our NbN film belongs to the dirty  regime with a Ioffe-Regel parameter $k_\mathrm{F} \ell=4.6$. Being in the dirty regime, we use the Werthamer-Helfand-Hohenberg relation~\cite{WerthamerPRB1966}, $B_\mathrm{c2}(0) = - 0.69 T_\mathrm{c} \frac{\mathrm{d}B_\mathrm{c2}}{\mathrm{d}T}$, to estimate the zero-temperature upper critical field $B_\mathrm{c2}(0)\approx$ \SI{18}{\tesla} (see inset of Fig.~\ref{fig:R(T)}). This implies a zero-temperature coherence length $\xi_0 = (\Phi_0/[2 \pi B_\mathrm{c2}(0)])^{1/2} = $ \SI{43}{\angstrom}, where $\Phi_0$ is the flux quantum. As our measurements are taken near the superconducting transition temperature where $\xi$ and the penetration depth~\cite{MondalPRL11a} diverge, the superconducting length scales are generally larger than the film thickness. Our system thus displays two-dimensional superconductivity whereas the electrons sense a three-dimensional environment due to their short mean-free path.

\begin{figure}
\includegraphics[width=\linewidth]{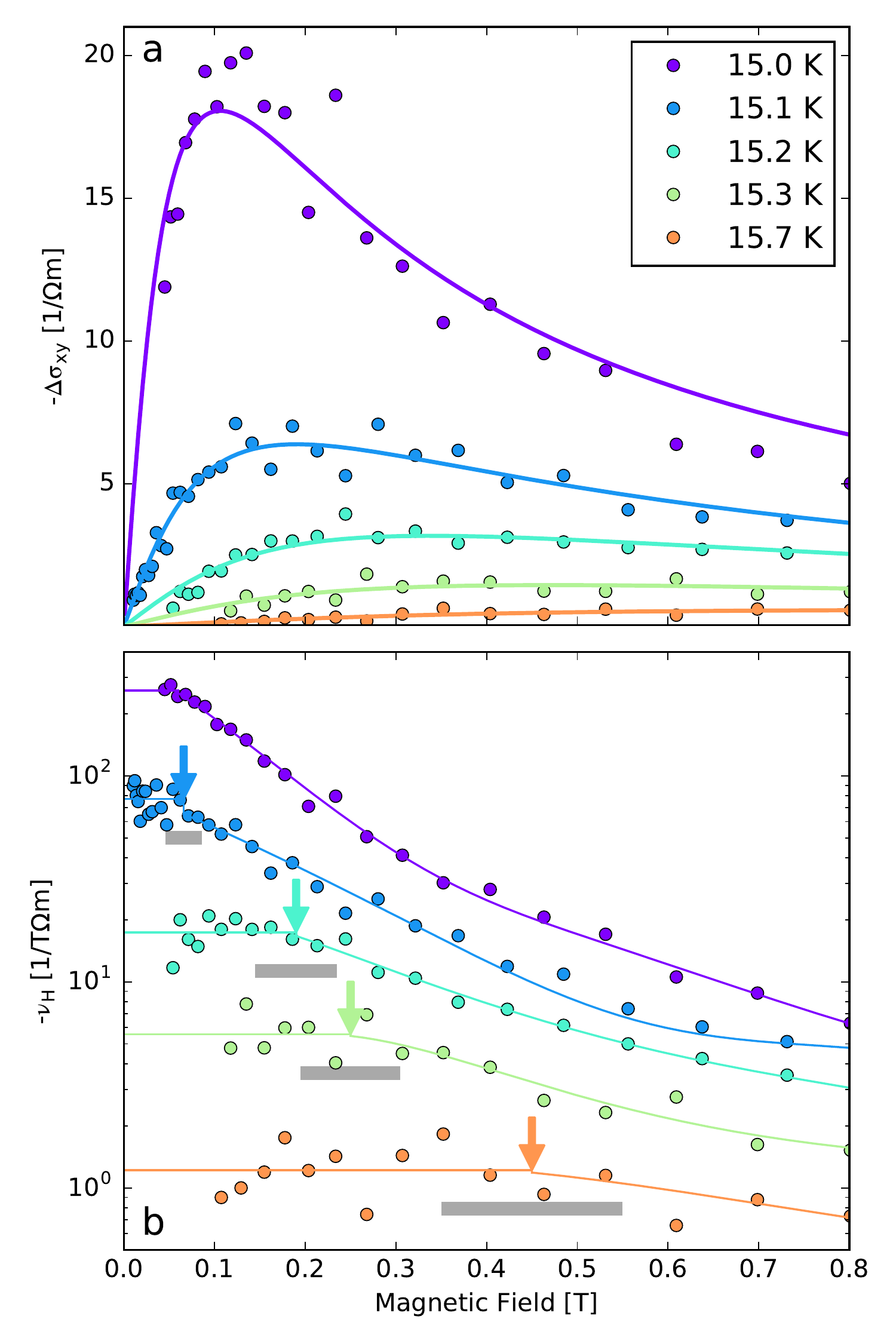}
\caption{\label{fig:delta_sigma}Hall conductivity due to superconducting fluctuations. The subtraction of the normal state response is described in the text. Isotherms of $\Delta \sigma_\mathrm{xy}$ shown in (a) are compared to Gaussian fluctuation theory (solid lines) explained in the text. The same data represented as $\nu_\mathrm{H}=\Delta \sigma_\mathrm{xy}/ B$ versus B are shown in (b).
For magnetic fields lower than $B^*$ (indicated by arrows), the isotherms of $\nu_\mathrm{H}$ become constant at values $\nu_\mathrm{H,0}$. For the isotherm at \SI{15.0}{\kelvin} the constant plateau is not reached at the lowest measurable fields and hence the flat line indicates a lower bound. The grey bars below the arrows show the estimated uncertainty of $B^*$ and all solid lines are guides to the eye.}
\end{figure}

\begin{figure}
\includegraphics[width=\linewidth]{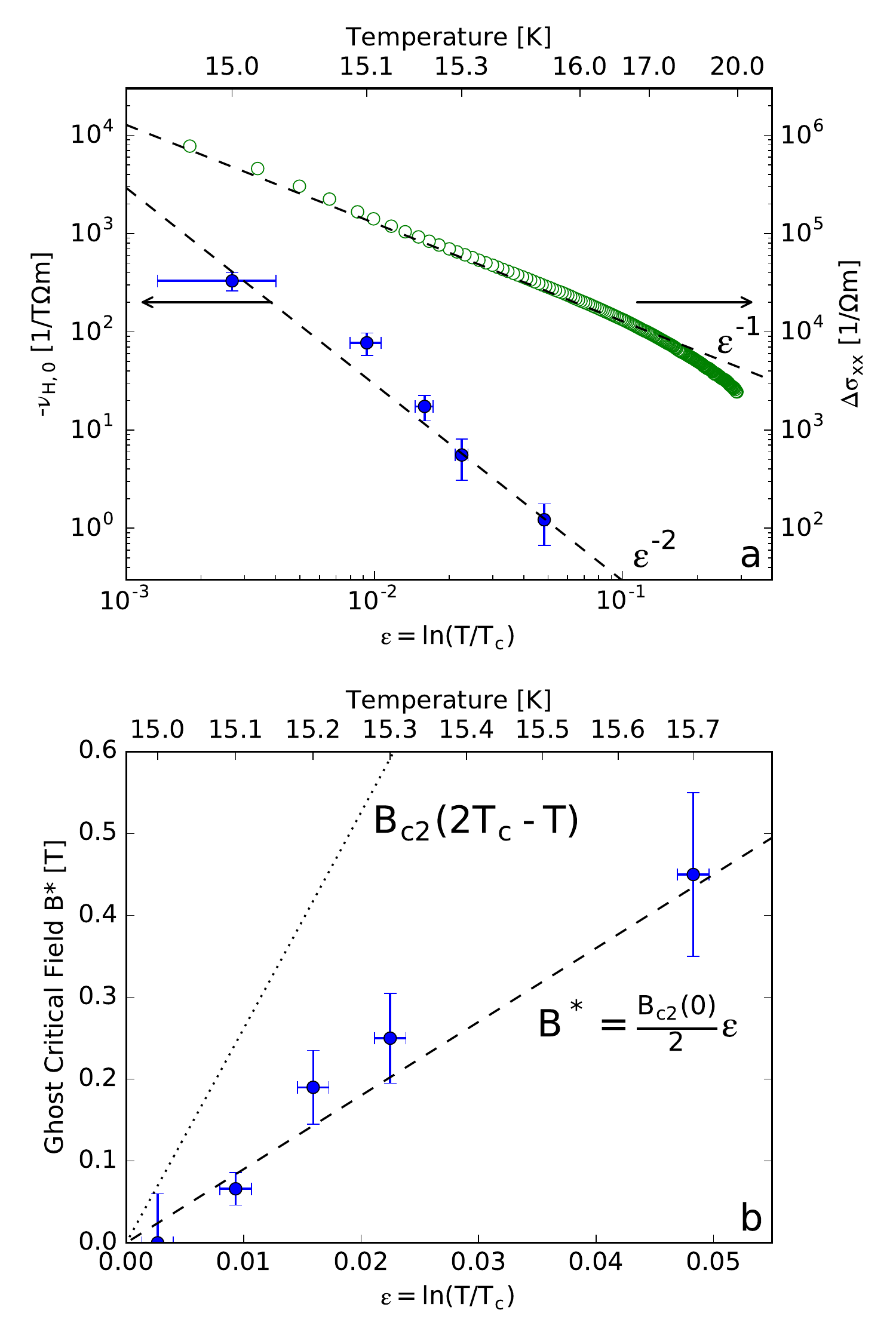}
\caption{\label{fig:slope_values_vs_epsilon}(a) The low-field value $\nu_\mathrm{H,0}=\Delta\sigma_\mathrm{xy}/B$ for ($B \rightarrow 0$) (left axis) and the paraconductivity $\Delta \sigma_\mathrm{xx}$ (right axis) as a function of $\epsilon$. Dashed lines are the predicted dependencies from Gaussian fluctuation theory without any adjustable parameters -- see text for a detailed explanation. (b) The ghost critical field as a function of $\epsilon$, obtained from the isotherms shown in Fig~\ref{fig:delta_sigma}. The dashed line corresponds to $B^*=B_\mathrm{c2}\epsilon/2$ with $B_\mathrm{c2}=18$~T obtained from the resistivity data shown in Fig.~\ref{fig:R(T)}. The dotted line corresponds to the mirror image of the red line in the inset of Fig.~\ref{fig:R(T)}. Vertical error bars correspond to the grey bars below the arrows in Fig.~\ref{fig:delta_sigma}. Horizontal error bars in (a) and (b) correspond to an uncertainty in $T_\mathrm{c}$ of \SI{20}{m\kelvin}.}
\end{figure}

Hall effect isotherms taken near the superconducting transition $T_\mathrm{c}$, display a sign change from negative to positive values at low magnetic fields (Fig.~\ref{fig:R_xy}a). This sign change is observed in a narrow temperature window of 0.3~K above $T_\mathrm{c}$. Deviations from linear low-field dependence is, however, observed up to $\sim1$~K above the superconducting transition. We thus analyse the isotherms in term of a negative normal state contribution $\rho_\mathrm{xy}\propto B$ and a positive response with a non-linear field dependence. To investigate the positive response, the negative linear normal state component is subtracted, i.e. $\Delta\rho_\mathrm{xy} = \rho_\mathrm{xy}-\rho_\mathrm{xy}^\mathrm{n}$. As shown in Fig.~\ref{fig:R_xy}b, the positive Hall effect response $\Delta \rho_\mathrm{xy}$  decreases rapidly with increasing temperature. In fact, it vanishes below the detection limit about 1~Kelvin above $T_\mathrm{c}$. 

Next, to compare with theoretical predictions, the contribution from superconductivity to the conductivity tensor is being isolated. As NbN displays essentially no magnetoresistance and $\rho_\mathrm{xx}\gg\rho_\mathrm{xy}$, the normal state conductivity $\sigma_\mathrm{xx}^\mathrm{n}$ is -- in the temperature regime of interest -- given by $\sigma_\mathrm{xx}^\mathrm{n} = 1 / \rho_\mathrm{xx} (\SI{9}{\tesla})$. The paraconductivity -- shown in Fig.~\ref{fig:slope_values_vs_epsilon}a --  is then given by $\Delta\sigma_\mathrm{xx} = \sigma_\mathrm{xx} - \sigma_\mathrm{xx}^\mathrm{n}$ where $\sigma_\mathrm{xx}=1 / \rho_\mathrm{xx} (\SI{0}{T})$. The Hall conductivity due to superconductivity is extracted in a similar fashion: $\Delta\sigma_\mathrm{xy}=\sigma_\mathrm{xy}-\sigma_\mathrm{xy}^\mathrm{n}$ where $\sigma_\mathrm{xy}^\mathrm{n} = - \rho_\mathrm{xy}^\mathrm{n} / [(\rho_\mathrm{xx}^\mathrm{n})^2 + (\rho_\mathrm{xy}^\mathrm{n})^2]$ and $\sigma_\mathrm{xy} = - \rho_\mathrm{xy} / [\rho_\mathrm{xx}^2 + \rho_\mathrm{xy}^2]$. In Fig.~\ref{fig:delta_sigma} isotherms of $-\Delta\sigma_\mathrm{xy}$ and $\nu_\mathrm{H}=\Delta\sigma_\mathrm{xy}/B$ for temperatures just above the superconducting transition $T_\mathrm{c}$ are shown. In the limit $B \rightarrow 0$, $\nu_\mathrm{H}(B)$ saturates at $\nu_\mathrm{H,0}$ and becomes essentially independent of magnetic field. The amplitude of the plateau ($\nu_\mathrm{H,0}$) is strongly temperature dependent. As shown in Fig.~\ref{fig:slope_values_vs_epsilon}a, $\nu_\mathrm{H,0}$ drops almost two orders of magnitude by heating just half a Kelvin above the superconducting transition. The onset of this low-field plateau defines a field scale $B^*$ that scales with $\epsilon = \ln(T/T_\mathrm{c})$ (Fig.~\ref{fig:slope_values_vs_epsilon}b).

\section{Discussion}
We now  discuss $\Delta\sigma_\mathrm{xx}$, $\Delta\sigma_\mathrm{xy}$, and the ghost critical field $B^*$. Generally, the paraconductivity $\Delta\sigma_\mathrm{xx}$ scales with the correlation length $\xi^2(T)$ that diverges as $T\rightarrow T_\mathrm{c}$. Gaussian fluctuations lead to a power-law divergence of the correlation length $\xi(T)\propto \epsilon^{-1/2}$. By contrast, phase fluctuations are expected to have an exponentially diverging correlation length $\xi(T)$. To bridge the two regimes, Halperin and Nelson~\cite{HalperinJLTP1979,MondalPRL11a} proposed a phenomenological function $\Delta \sigma_\mathrm{xx} \propto  \sinh^2{(\sqrt{b \tau_\mathrm{c}/ \tau})}$ -- where $b$ is a constant, $\tau_\mathrm{c} = (T_\mathrm{c}-T_\mathrm{KT})/T_\mathrm{KT}$, and $\tau=(T-T_\mathrm{KT})/T_\mathrm{KT}$. Phase fluctuating superconductivity would exist in  between $T_\mathrm{KT}$ -- the Kosterlitz-Thouless transition temperature -- and $T_\mathrm{c}$.  For $b\tau_\mathrm{c}/\tau\gg 1$, an exponential divergence $\Delta\sigma_\mathrm{xx} \propto \exp(2\sqrt{b\tau_\mathrm{c}/\tau})$ is dominant whereas for $b\tau_\mathrm{c}/\tau\ll 1$ the Gaussian power-law dependence is recovered. It has been demonstrated theoretically that the dimensionless constant $b$ scales with $\sqrt{\tau_\mathrm{c}}$~\cite{BenfattoPRB2009}. As the superconducting transition of our film is very sharp, it is clear that $\tau_\mathrm{c}$ and hence $b$ are vanishing small. The Kosterlitz-Thouless regime is thus not relevant for our system and we therefore discuss the superconducting fluctuations  within the Gaussian paradigm.  

A central prediction is that the normal state superconducting fluctuations should display a characteristic field scale $B^*$ that in essence marks the cross-over between coherence length $\xi(T)=\xi/\sqrt{\epsilon}$ and the magnetic length scale $\ell_B=\sqrt{\hbar/\alpha eB}$. Different numerical values of $\alpha=1,2$, and 4 have previously been used~\cite{BreznayPRB2012,BreznayPRB13,PourretPRB07,PourretNatPhys2006,PourretNJP09}.  The cross-over condition ($\xi(T)=\ell_B$) implies that $B^*= 2 B_\mathrm{c2}(0)\epsilon /\alpha$ and since $B^*$ scales with $B_\mathrm{c2}(0)$, it is often referred to as the ghost critical field scale. For Nernst effect isotherms ($N$ versus $B$), this field scale is commonly defined by a maximum in $N$~\cite{TaftiPRB14,PourretNatPhys2006,ChangNatPhys2012}. However, this maximum is not expected to strictly vanish for $T\rightarrow T_\mathrm{c}$ and indeed Nernst effect experiments typically find a saturation of $B^*$ near $T_\mathrm{c}$~\cite{ChangNatPhys2012, Yamashita2015}. For this reason, it makes sense to define the  ghost critical field $B^*$ as the field scale below which $\nu=N/B$ or $\nu_\mathrm{H}=\Delta\sigma_\mathrm{xy}/B$ is constant (see Fig.~\ref{fig:delta_sigma}). Within error bars the extracted ghost critical field scales with $\epsilon$ as shown in Fig.~\ref{fig:slope_values_vs_epsilon}b. Moreover, using $B_\mathrm{c2}(0) \approx$~\SI{18}{\tesla} (derived from the resistivity curves -- shown in Fig.~\ref{fig:R(T)}) good agreement with the experiment is obtained with $\alpha=4$. This implies, as has previously been observed in TaN and cuprate superconductors~\cite{TaftiPRB14,BreznayPRB2012,BreznayPRB13}, that the ghost critical field $B^*$ is generally lower than the upper critical field $B_\mathrm{c2}(0)$. 

Gaussian / amplitude fluctuations of superconductivity constitute short-lived Cooper pairs that open a new channel for charge transport. In two dimensions, this yields the following contributions to the conductivity tensor~\cite{TikhonovPRB12,MichaeliPRB2012}:
\begin{equation}\label{eq:eq2}
\Delta\sigma_\mathrm{xx}^{(1)}=\frac{e^2}{16\hbar d}\frac{1}{\epsilon} \quad \textrm{and} \quad \frac{\Delta\sigma_\mathrm{xy}^{(1)}}{B}=\frac{ |e| D \kappa}{3}\frac{e^2}{16\hbar d}\frac{1}{\epsilon^2}
\end{equation}
where $\Delta\sigma_\mathrm{xx}$ (the Aslamazov-Larkin (AL) term \cite{Aslamazov1968}) is independent of material properties and $\Delta\sigma_\mathrm{xy}$ depends only on the diffusion constant $D$ and $\kappa$. The AL-term is expected valid in the regime $Gi < \epsilon \ll 1$~\cite{VarlamovBook}, where $Gi= (e^2 \rho^n_\mathrm{xx})/(16 \hbar d) \approx 0.001$ is the Ginzburg-Levanyuk parameter here defined by the condition $\Delta\sigma_\mathrm{xx}^{(1)}=\sigma^n_{xx}$~\cite{VarlamovBook}. Scattering of electrons on the fluctuating superconductivity is described by the so-called Maki-Thompson terms~\cite{Maki68,ThompsonPRB70,LangPRB94}:
\begin{equation}\label{eq:eq3}
\Delta\sigma_\mathrm{xx}^{(2)}=  \frac{2\epsilon \Delta\sigma_\mathrm{xx}^{(1)}}{\epsilon-\delta} \ln\left(\frac{\epsilon}{\delta}\right)  \quad \textrm{and} \quad \frac{\Delta\sigma_\mathrm{xy}^{(2)}}{B}=-2\mu_\mathrm{H}\Delta\sigma_\mathrm{xx}^{(2)}
\end{equation}
where $\delta=\pi \tau_0 / (8\tau_\phi)$, $\tau_0=\hbar/k_\mathrm{B} T$, $\tau_\phi$ is the dephasing time~\cite{ReizerPRB92}, and $\mu_\mathrm{H}$ is the electron mobility. We have no direct experimental measure of the dephasing time $\tau_\phi$ in NbN. For the optimally doped high-temperature superconductor YBCO, $\tau_\phi=$ \SI{86}{\femto\s} has previously been reported~\cite{LangPRB94}. Assuming a similar or shorter dephasing time for NbN, implies that $\Delta\sigma_\mathrm{xx}^{(1)} \gg \Delta\sigma_\mathrm{xx}^{(2)}$ and  $\Delta\sigma_\mathrm{xy}^{(1)} \gg \Delta\sigma_\mathrm{xy}^{(2)}$. The absence of a $\ln(\epsilon/\delta)$ dependence of $\Delta\sigma_\mathrm{xx}$ (Fig.~\ref{fig:slope_values_vs_epsilon}a) lends experimental support for this ansatz. Finally, density-of-state (DOS) corrections are predicted to be subleading contributions~\cite{TikhonovPRB12}. We therefore discuss the experimentally observed paraconductivity by setting $\Delta \sigma_i=\Delta \sigma_i^{(1)}$ with $i \in \{\mathrm{xx},\mathrm{xy}\}$.

The exact experimental $T$-dependence of $\Delta \sigma_i$ in the limit $\epsilon\rightarrow 0$ is sensitive to how $T_\mathrm{c}$ is defined~\cite{LeridonPRB2007,BreznayPRB2012}. When defining $T_\mathrm{c}=$~\SI{14.96}{\kelvin} by the steepest point of the superconducting transition, we find that $\Delta\sigma_\mathrm{xx}$ scales perfectly with $\epsilon^{-1}$ and $-\nu_\mathrm{H,0}$ is consistent with a $\epsilon^{-2}$ dependence (Fig.~\ref{fig:slope_values_vs_epsilon}a). Gaussian fluctuation theory thus provides an excellent description of the temperature dependence of $\Delta\sigma_\mathrm{xx}$, $\Delta\sigma_\mathrm{xy}$ and $B^*$ without any adjustment of $T_\mathrm{c}$.

The absolute magnitude of $\Delta\sigma_\mathrm{xx}$ and $\Delta\sigma_\mathrm{xy}$ is also well described by Gaussian fluctuation theory. Using the film thickness $d=119$~\AA, the  predicted magnitude of $\Delta\sigma_\mathrm{xx}$ is within 20$\%$ in agreement with the experiment (Fig.~\ref{fig:slope_values_vs_epsilon}a). The amplitude of  $\Delta\sigma_\mathrm{xy}$ can be evaluated by estimating the product $|e|D\kappa$. Notice that $\kappa=\beta/\epsilon_\mathrm{F}$ where $\epsilon_\mathrm{F}$ is the Fermi energy and $\beta=-0.5\lambda \cdot \mathrm{d}\ln(T_\mathrm{c})/\mathrm{d}\lambda$ is depending on the superconducting coupling constant $\lambda$~\cite{MichaeliPRB2012}. In the weak coupling limit ($\lambda\ll 1$), where $T_\mathrm{c}\propto \exp(-1/\lambda)$~\cite{BardeenPR1957} one finds $\beta=-1/(2\lambda)$. NbN, however, belongs to the strongly coupled $\lambda\sim 1$ limit~\cite{ChockalingamPRB2008,BeckPRL11} where 
$T_\mathrm{c} \propto \exp([-1.04(1+\lambda)]/[\lambda-\mu^*(1+0.62\lambda)])$
with $\mu^*=0.13$ being the screening potential~\cite{McMillanPR68,ChockalingamPRB2008}. A recent experimental study found $\lambda=1.1\pm 0.1$~\cite{BeckPRL11} implying that $\beta=-0.77$. Finally, using $D=v_\mathrm{F}^2\tau/3$ where $v_\mathrm{F}$ is the Fermi velocity and $\tau$ is the mean free life time, it is found that $|e|D\kappa=2\mu_\mathrm{H}\beta/3$ where $\mu_\mathrm{H}=|e|\tau/m^*$ is the Hall mobility and $m^*$ is the quasiparticle mass. The Hall mobility can be derived directly from the experiment  $\mu_\mathrm{H}=|R_\mathrm{H}/\rho_\mathrm{xx}^\mathrm{n}|=$\SI{1.3e-5}{\tesla ^{-1}}. We find, without any adjustable parameters, that the theoretical prediction of $\Delta\sigma_\mathrm{xy}$ (solid line in Fig.~\ref{fig:slope_values_vs_epsilon}a) is in excellent agreement with the experiment.

Finally, it is also possible -- as done  in Ref.~\onlinecite{BreznayPRB2012} -- to analyze a more extended region of magnetic fields by fitting the isotherms. 
 Gaussian theory predicts the isotherms to be described by
\begin{equation}
\Delta\sigma_\mathrm{xy}^{(1)}(\omega\rightarrow0)=\frac{e^2k_\mathrm{B}T\kappa}{\pi\hbar d} \mathrm{sgn}(B) \sum_{j=0}^\infty \frac{(j+1)(\zeta_{j+1}-\zeta_j )^3}{\zeta_{j}\zeta_{j+1}(\zeta_{j+1}+\zeta_j )^2}
\end{equation}
where 
\begin{equation}
\zeta_j=\epsilon + \frac{\kappa\omega}{2} + \Psi\left(\frac{1}{2}+\frac{(j+1/2)4D|eB|-i\omega}{4\pi k_\mathrm{B} T}\right)-\Psi\left(\frac{1}{2}\right)
\end{equation}
and  $\Psi$ is the digamma function. By using $\kappa$ and $D$ as fit parameters it is possible, as shown in Fig.~\ref{fig:delta_sigma}a, to describe reasonably well the isotherms of $\sigma_\mathrm{xy}$. The obtained values of $\kappa$ and $D$ are in good agreement with the $B\rightarrow 0$ analysis presented in Fig.~\ref{fig:slope_values_vs_epsilon}a.  As a concluding remark, we stress the advantage of the  $B\rightarrow 0$ analysis. As the product $\kappa D$  is proportional to the Hall mobility, the Gaussian theory can be tested without knowing or fitting $\kappa$ and $D$.

\section{Conclusions}
In summary, we presented a systematic study of the conductivity response generated by superconducting fluctuations in the normal state of a \SI{119}{\angstrom} thin NbN film. It is shown how these fluctuations drive a sign change in the Hall coefficient. Isolating the longitudinal $\Delta\sigma_\mathrm{xx}$ and transverse $\Delta\sigma_\mathrm{xy}$ conductivity due to superconducting fluctuations allowed direct comparison to Gaussian fluctuation theory. We found that these transport quantities are scaling with the distance to the superconducting transition $\epsilon=\ln(T/T_\mathrm{c})$ as predicted ($\Delta\sigma_\mathrm{xx}\propto\epsilon^{-1}$ and $\Delta\sigma_\mathrm{xy}\propto\epsilon^{-2}$). Furthermore, excellent quantitative agreement between Gaussian fluctuation theory and the experiment was obtained. The presented study thus demonstrates experimentally how Gaussian fluctuations of superconductivity contribute to the conductivity tensor.

\begin{acknowledgments}
The authors thank X. Zhang and K. Michaeli for useful discussions. D.~D. and J.~C. acknowledge support by the Swiss National Science Foundation under grant PP00P2 \textunderscore 150573.
\end{acknowledgments}


\begin{thebibliography}{48}%
\makeatletter
\providecommand \@ifxundefined [1]{%
 \@ifx{#1\undefined}
}%
\providecommand \@ifnum [1]{%
 \ifnum #1\expandafter \@firstoftwo
 \else \expandafter \@secondoftwo
 \fi
}%
\providecommand \@ifx [1]{%
 \ifx #1\expandafter \@firstoftwo
 \else \expandafter \@secondoftwo
 \fi
}%
\providecommand \natexlab [1]{#1}%
\providecommand \enquote  [1]{``#1''}%
\providecommand \bibnamefont  [1]{#1}%
\providecommand \bibfnamefont [1]{#1}%
\providecommand \citenamefont [1]{#1}%
\providecommand \href@noop [0]{\@secondoftwo}%
\providecommand \href [0]{\begingroup \@sanitize@url \@href}%
\providecommand \@href[1]{\@@startlink{#1}\@@href}%
\providecommand \@@href[1]{\endgroup#1\@@endlink}%
\providecommand \@sanitize@url [0]{\catcode `\\12\catcode `\$12\catcode
  `\&12\catcode `\#12\catcode `\^12\catcode `\_12\catcode `\%12\relax}%
\providecommand \@@startlink[1]{}%
\providecommand \@@endlink[0]{}%
\providecommand \url  [0]{\begingroup\@sanitize@url \@url }%
\providecommand \@url [1]{\endgroup\@href {#1}{\urlprefix }}%
\providecommand \urlprefix  [0]{URL }%
\providecommand \Eprint [0]{\href }%
\providecommand \doibase [0]{http://dx.doi.org/}%
\providecommand \selectlanguage [0]{\@gobble}%
\providecommand \bibinfo  [0]{\@secondoftwo}%
\providecommand \bibfield  [0]{\@secondoftwo}%
\providecommand \translation [1]{[#1]}%
\providecommand \BibitemOpen [0]{}%
\providecommand \bibitemStop [0]{}%
\providecommand \bibitemNoStop [0]{.\EOS\space}%
\providecommand \EOS [0]{\spacefactor3000\relax}%
\providecommand \BibitemShut  [1]{\csname bibitem#1\endcsname}%
\let\auto@bib@innerbib\@empty
\bibitem [{\citenamefont {Ussishkin}\ \emph {et~al.}(2002)\citenamefont
  {Ussishkin}, \citenamefont {Sondhi},\ and\ \citenamefont
  {Huse}}]{UssishkinPRL02}%
  \BibitemOpen
  \bibfield  {author} {\bibinfo {author} {\bibfnamefont {I.}~\bibnamefont
  {Ussishkin}}, \bibinfo {author} {\bibfnamefont {S.~L.}\ \bibnamefont
  {Sondhi}}, \ and\ \bibinfo {author} {\bibfnamefont {D.~A.}\ \bibnamefont
  {Huse}},\ }\href {\doibase 10.1103/PhysRevLett.89.287001} {\bibfield
  {journal} {\bibinfo  {journal} {Phys. Rev. Lett.}\ }\textbf {\bibinfo
  {volume} {89}},\ \bibinfo {pages} {287001} (\bibinfo {year}
  {2002})}\BibitemShut {NoStop}%
\bibitem [{\citenamefont {Emery}\ and\ \citenamefont
  {Kivelson}(1995)}]{EmeryNature1995}%
  \BibitemOpen
  \bibfield  {author} {\bibinfo {author} {\bibfnamefont {V.~J.}\ \bibnamefont
  {Emery}}\ and\ \bibinfo {author} {\bibfnamefont {S.~A.}\ \bibnamefont
  {Kivelson}},\ }\href {\doibase 10.1038/374434a0} {\bibfield  {journal}
  {\bibinfo  {journal} {Nature}\ }\textbf {\bibinfo {volume} {374}},\ \bibinfo
  {pages} {434} (\bibinfo {year} {1995})}\BibitemShut {NoStop}%
\bibitem [{\citenamefont {Randeria}(2010)}]{RanderiaNatPhys2010}%
  \BibitemOpen
  \bibfield  {author} {\bibinfo {author} {\bibfnamefont {M.}~\bibnamefont
  {Randeria}},\ }\href {\doibase 10.1038/nphys1748} {\bibfield  {journal}
  {\bibinfo  {journal} {Nat. Phys.}\ }\textbf {\bibinfo {volume} {6}},\
  \bibinfo {pages} {561} (\bibinfo {year} {2010})}\BibitemShut {NoStop}%
\bibitem [{\citenamefont {Gantmakher}\ and\ \citenamefont
  {Dolgopolov}(2010)}]{GantmakherPU10}%
  \BibitemOpen
  \bibfield  {author} {\bibinfo {author} {\bibfnamefont {V.~F.}\ \bibnamefont
  {Gantmakher}}\ and\ \bibinfo {author} {\bibfnamefont {V.~T.}\ \bibnamefont
  {Dolgopolov}},\ }\href {http://stacks.iop.org/1063-7869/53/i=1/a=R01}
  {\bibfield  {journal} {\bibinfo  {journal} {Physics-Uspekhi}\ }\textbf
  {\bibinfo {volume} {53}},\ \bibinfo {pages} {1} (\bibinfo {year}
  {2010})}\BibitemShut {NoStop}%
\bibitem [{\citenamefont {Michaeli}\ and\ \citenamefont
  {Finkel'stein}(2009)}]{MichaeliEPL09}%
  \BibitemOpen
  \bibfield  {author} {\bibinfo {author} {\bibfnamefont {K.}~\bibnamefont
  {Michaeli}}\ and\ \bibinfo {author} {\bibfnamefont {A.~M.}\ \bibnamefont
  {Finkel'stein}},\ }\href {http://stacks.iop.org/0295-5075/86/i=2/a=27007}
  {\bibfield  {journal} {\bibinfo  {journal} {EPL (Europhysics Letters)}\
  }\textbf {\bibinfo {volume} {86}},\ \bibinfo {pages} {27007} (\bibinfo {year}
  {2009})}\BibitemShut {NoStop}%
\bibitem [{\citenamefont {Serbyn}\ \emph {et~al.}(2009)\citenamefont {Serbyn},
  \citenamefont {Skvortsov}, \citenamefont {Varlamov},\ and\ \citenamefont
  {Galitski}}]{SerbynPRL09}%
  \BibitemOpen
  \bibfield  {author} {\bibinfo {author} {\bibfnamefont {M.~N.}\ \bibnamefont
  {Serbyn}}, \bibinfo {author} {\bibfnamefont {M.~A.}\ \bibnamefont
  {Skvortsov}}, \bibinfo {author} {\bibfnamefont {A.~A.}\ \bibnamefont
  {Varlamov}}, \ and\ \bibinfo {author} {\bibfnamefont {V.}~\bibnamefont
  {Galitski}},\ }\href {\doibase 10.1103/PhysRevLett.102.067001} {\bibfield
  {journal} {\bibinfo  {journal} {Phys. Rev. Lett.}\ }\textbf {\bibinfo
  {volume} {102}},\ \bibinfo {pages} {067001} (\bibinfo {year}
  {2009})}\BibitemShut {NoStop}%
\bibitem [{\citenamefont {Xu}\ \emph {et~al.}(2000)\citenamefont {Xu},
  \citenamefont {Ong}, \citenamefont {Wang}, \citenamefont {Kakeshita},\ and\
  \citenamefont {Uchida}}]{XuNat00}%
  \BibitemOpen
  \bibfield  {author} {\bibinfo {author} {\bibfnamefont {Z.~A.}\ \bibnamefont
  {Xu}}, \bibinfo {author} {\bibfnamefont {N.~P.}\ \bibnamefont {Ong}},
  \bibinfo {author} {\bibfnamefont {Y.}~\bibnamefont {Wang}}, \bibinfo {author}
  {\bibfnamefont {T.}~\bibnamefont {Kakeshita}}, \ and\ \bibinfo {author}
  {\bibfnamefont {S.}~\bibnamefont {Uchida}},\ }\href {\doibase
  10.1038/35020016} {\bibfield  {journal} {\bibinfo  {journal} {Nature}\
  }\textbf {\bibinfo {volume} {406}},\ \bibinfo {pages} {486} (\bibinfo {year}
  {2000})}\BibitemShut {NoStop}%
\bibitem [{\citenamefont {Wang}\ \emph {et~al.}(2006)\citenamefont {Wang},
  \citenamefont {Li},\ and\ \citenamefont {Ong}}]{WangPRB03}%
  \BibitemOpen
  \bibfield  {author} {\bibinfo {author} {\bibfnamefont {Y.}~\bibnamefont
  {Wang}}, \bibinfo {author} {\bibfnamefont {L.}~\bibnamefont {Li}}, \ and\
  \bibinfo {author} {\bibfnamefont {N.~P.}\ \bibnamefont {Ong}},\ }\href
  {\doibase 10.1103/PhysRevB.73.024510} {\bibfield  {journal} {\bibinfo
  {journal} {Phys. Rev. B}\ }\textbf {\bibinfo {volume} {73}},\ \bibinfo
  {pages} {024510} (\bibinfo {year} {2006})}\BibitemShut {NoStop}%
\bibitem [{\citenamefont {Li}\ \emph {et~al.}(2010)\citenamefont {Li},
  \citenamefont {Wang}, \citenamefont {Komiya}, \citenamefont {Ono},
  \citenamefont {Ando}, \citenamefont {Gu},\ and\ \citenamefont
  {Ong}}]{LiPRB2010}%
  \BibitemOpen
  \bibfield  {author} {\bibinfo {author} {\bibfnamefont {L.}~\bibnamefont
  {Li}}, \bibinfo {author} {\bibfnamefont {Y.}~\bibnamefont {Wang}}, \bibinfo
  {author} {\bibfnamefont {S.}~\bibnamefont {Komiya}}, \bibinfo {author}
  {\bibfnamefont {S.}~\bibnamefont {Ono}}, \bibinfo {author} {\bibfnamefont
  {Y.}~\bibnamefont {Ando}}, \bibinfo {author} {\bibfnamefont {G.~D.}\
  \bibnamefont {Gu}}, \ and\ \bibinfo {author} {\bibfnamefont {N.~P.}\
  \bibnamefont {Ong}},\ }\href {\doibase 10.1103/PhysRevB.81.054510} {\bibfield
   {journal} {\bibinfo  {journal} {Phys. Rev. B}\ }\textbf {\bibinfo {volume}
  {81}},\ \bibinfo {pages} {054510} (\bibinfo {year} {2010})}\BibitemShut
  {NoStop}%
\bibitem [{\citenamefont {Baity}\ \emph {et~al.}(2016)\citenamefont {Baity},
  \citenamefont {Shi}, \citenamefont {Shi}, \citenamefont {Benfatto},\ and\
  \citenamefont {Popovi\ifmmode~\acute{c}\else \'{c}\fi{}}}]{BaityPRB2016}%
  \BibitemOpen
  \bibfield  {author} {\bibinfo {author} {\bibfnamefont {P.~G.}\ \bibnamefont
  {Baity}}, \bibinfo {author} {\bibfnamefont {X.}~\bibnamefont {Shi}}, \bibinfo
  {author} {\bibfnamefont {Z.}~\bibnamefont {Shi}}, \bibinfo {author}
  {\bibfnamefont {L.}~\bibnamefont {Benfatto}}, \ and\ \bibinfo {author}
  {\bibfnamefont {D.}~\bibnamefont {Popovi\ifmmode~\acute{c}\else
  \'{c}\fi{}}},\ }\href {\doibase 10.1103/PhysRevB.93.024519} {\bibfield
  {journal} {\bibinfo  {journal} {Phys. Rev. B}\ }\textbf {\bibinfo {volume}
  {93}},\ \bibinfo {pages} {024519} (\bibinfo {year} {2016})}\BibitemShut
  {NoStop}%
\bibitem [{\citenamefont {Chang}\ \emph {et~al.}(2012)\citenamefont {Chang},
  \citenamefont {Doiron-Leyraud}, \citenamefont {Cyr-Choiniere}, \citenamefont
  {Grissonnanche}, \citenamefont {Laliberte}, \citenamefont {Hassinger},
  \citenamefont {Reid}, \citenamefont {Daou}, \citenamefont {Pyon},
  \citenamefont {Takayama}, \citenamefont {Takagi},\ and\ \citenamefont
  {Taillefer}}]{ChangNatPhys2012}%
  \BibitemOpen
  \bibfield  {author} {\bibinfo {author} {\bibfnamefont {J.}~\bibnamefont
  {Chang}}, \bibinfo {author} {\bibfnamefont {N.}~\bibnamefont
  {Doiron-Leyraud}}, \bibinfo {author} {\bibfnamefont {O.}~\bibnamefont
  {Cyr-Choiniere}}, \bibinfo {author} {\bibfnamefont {G.}~\bibnamefont
  {Grissonnanche}}, \bibinfo {author} {\bibfnamefont {F.}~\bibnamefont
  {Laliberte}}, \bibinfo {author} {\bibfnamefont {E.}~\bibnamefont
  {Hassinger}}, \bibinfo {author} {\bibfnamefont {J.-P.}\ \bibnamefont {Reid}},
  \bibinfo {author} {\bibfnamefont {R.}~\bibnamefont {Daou}}, \bibinfo {author}
  {\bibfnamefont {S.}~\bibnamefont {Pyon}}, \bibinfo {author} {\bibfnamefont
  {T.}~\bibnamefont {Takayama}}, \bibinfo {author} {\bibfnamefont
  {H.}~\bibnamefont {Takagi}}, \ and\ \bibinfo {author} {\bibfnamefont
  {L.}~\bibnamefont {Taillefer}},\ }\href {\doibase 10.1038/nphys2380}
  {\bibfield  {journal} {\bibinfo  {journal} {Nat. Phys.}\ }\textbf {\bibinfo
  {volume} {8}},\ \bibinfo {pages} {751} (\bibinfo {year} {2012})}\BibitemShut
  {NoStop}%
\bibitem [{\citenamefont {Pourret}\ \emph {et~al.}(2006)\citenamefont
  {Pourret}, \citenamefont {Aubin}, \citenamefont {Lesueur}, \citenamefont
  {Marrache-Kikuchi}, \citenamefont {Berge}, \citenamefont {Dumoulin},\ and\
  \citenamefont {Behnia}}]{PourretNatPhys2006}%
  \BibitemOpen
  \bibfield  {author} {\bibinfo {author} {\bibfnamefont {A.}~\bibnamefont
  {Pourret}}, \bibinfo {author} {\bibfnamefont {H.}~\bibnamefont {Aubin}},
  \bibinfo {author} {\bibfnamefont {J.}~\bibnamefont {Lesueur}}, \bibinfo
  {author} {\bibfnamefont {C.~A.}\ \bibnamefont {Marrache-Kikuchi}}, \bibinfo
  {author} {\bibfnamefont {L.}~\bibnamefont {Berge}}, \bibinfo {author}
  {\bibfnamefont {L.}~\bibnamefont {Dumoulin}}, \ and\ \bibinfo {author}
  {\bibfnamefont {K.}~\bibnamefont {Behnia}},\ }\href {\doibase
  10.1038/nphys413} {\bibfield  {journal} {\bibinfo  {journal} {Nat. Phys.}\
  }\textbf {\bibinfo {volume} {2}},\ \bibinfo {pages} {683} (\bibinfo {year}
  {2006})}\BibitemShut {NoStop}%
\bibitem [{\citenamefont {Cabo}\ \emph {et~al.}(2007)\citenamefont {Cabo},
  \citenamefont {Mosqueira},\ and\ \citenamefont {Vidal}}]{CaboPRL07}%
  \BibitemOpen
  \bibfield  {author} {\bibinfo {author} {\bibfnamefont {L.}~\bibnamefont
  {Cabo}}, \bibinfo {author} {\bibfnamefont {J.}~\bibnamefont {Mosqueira}}, \
  and\ \bibinfo {author} {\bibfnamefont {F.}~\bibnamefont {Vidal}},\ }\href
  {\doibase 10.1103/PhysRevLett.98.119701} {\bibfield  {journal} {\bibinfo
  {journal} {Phys. Rev. Lett.}\ }\textbf {\bibinfo {volume} {98}},\ \bibinfo
  {pages} {119701} (\bibinfo {year} {2007})}\BibitemShut {NoStop}%
\bibitem [{\citenamefont {Rullier-Albenque}\ \emph {et~al.}(2011)\citenamefont
  {Rullier-Albenque}, \citenamefont {Alloul},\ and\ \citenamefont
  {Rikken}}]{RullierAlbenquePRB2011}%
  \BibitemOpen
  \bibfield  {author} {\bibinfo {author} {\bibfnamefont {F.}~\bibnamefont
  {Rullier-Albenque}}, \bibinfo {author} {\bibfnamefont {H.}~\bibnamefont
  {Alloul}}, \ and\ \bibinfo {author} {\bibfnamefont {G.}~\bibnamefont
  {Rikken}},\ }\href {\doibase 10.1103/PhysRevB.84.014522} {\bibfield
  {journal} {\bibinfo  {journal} {Phys. Rev. B}\ }\textbf {\bibinfo {volume}
  {84}},\ \bibinfo {pages} {014522} (\bibinfo {year} {2011})}\BibitemShut
  {NoStop}%
\bibitem [{\citenamefont {Leridon}\ \emph {et~al.}(2007)\citenamefont
  {Leridon}, \citenamefont {Vanacken}, \citenamefont {Wambecq},\ and\
  \citenamefont {Moshchalkov}}]{LeridonPRB2007}%
  \BibitemOpen
  \bibfield  {author} {\bibinfo {author} {\bibfnamefont {B.}~\bibnamefont
  {Leridon}}, \bibinfo {author} {\bibfnamefont {J.}~\bibnamefont {Vanacken}},
  \bibinfo {author} {\bibfnamefont {T.}~\bibnamefont {Wambecq}}, \ and\
  \bibinfo {author} {\bibfnamefont {V.~V.}\ \bibnamefont {Moshchalkov}},\
  }\href {\doibase 10.1103/PhysRevB.76.012503} {\bibfield  {journal} {\bibinfo
  {journal} {Phys. Rev. B}\ }\textbf {\bibinfo {volume} {76}},\ \bibinfo
  {pages} {012503} (\bibinfo {year} {2007})}\BibitemShut {NoStop}%
\bibitem [{\citenamefont {Perfetti}\ \emph {et~al.}(2015)\citenamefont
  {Perfetti}, \citenamefont {Sciolla}, \citenamefont {Biroli}, \citenamefont
  {van~der Beek}, \citenamefont {Piovera}, \citenamefont {Wolf},\ and\
  \citenamefont {Kampfrath}}]{PerfettiPRL15}%
  \BibitemOpen
  \bibfield  {author} {\bibinfo {author} {\bibfnamefont {L.}~\bibnamefont
  {Perfetti}}, \bibinfo {author} {\bibfnamefont {B.}~\bibnamefont {Sciolla}},
  \bibinfo {author} {\bibfnamefont {G.}~\bibnamefont {Biroli}}, \bibinfo
  {author} {\bibfnamefont {C.~J.}\ \bibnamefont {van~der Beek}}, \bibinfo
  {author} {\bibfnamefont {C.}~\bibnamefont {Piovera}}, \bibinfo {author}
  {\bibfnamefont {M.}~\bibnamefont {Wolf}}, \ and\ \bibinfo {author}
  {\bibfnamefont {T.}~\bibnamefont {Kampfrath}},\ }\href {\doibase
  10.1103/PhysRevLett.114.067003} {\bibfield  {journal} {\bibinfo  {journal}
  {Phys. Rev. Lett.}\ }\textbf {\bibinfo {volume} {114}},\ \bibinfo {pages}
  {067003} (\bibinfo {year} {2015})}\BibitemShut {NoStop}%
\bibitem [{\citenamefont {Behnia}\ and\ \citenamefont
  {Aubin}(2016)}]{BehniaRPP16}%
  \BibitemOpen
  \bibfield  {author} {\bibinfo {author} {\bibfnamefont {K.}~\bibnamefont
  {Behnia}}\ and\ \bibinfo {author} {\bibfnamefont {H.}~\bibnamefont {Aubin}},\
  }\href {http://stacks.iop.org/0034-4885/79/i=4/a=046502} {\bibfield
  {journal} {\bibinfo  {journal} {Reports on Progress in Physics}\ }\textbf
  {\bibinfo {volume} {79}},\ \bibinfo {pages} {046502} (\bibinfo {year}
  {2016})}\BibitemShut {NoStop}%
\bibitem [{\citenamefont {Sac{\'e}p{\'e}}\ \emph {et~al.}(2010)\citenamefont
  {Sac{\'e}p{\'e}}, \citenamefont {Chapelier}, \citenamefont {Baturina},
  \citenamefont {Vinokur}, \citenamefont {Baklanov},\ and\ \citenamefont
  {Sanquer}}]{SacepeNATCOMM10}%
  \BibitemOpen
  \bibfield  {author} {\bibinfo {author} {\bibfnamefont {B.}~\bibnamefont
  {Sac{\'e}p{\'e}}}, \bibinfo {author} {\bibfnamefont {C.}~\bibnamefont
  {Chapelier}}, \bibinfo {author} {\bibfnamefont {T.~I.}\ \bibnamefont
  {Baturina}}, \bibinfo {author} {\bibfnamefont {V.~M.}\ \bibnamefont
  {Vinokur}}, \bibinfo {author} {\bibfnamefont {M.~R.}\ \bibnamefont
  {Baklanov}}, \ and\ \bibinfo {author} {\bibfnamefont {M.}~\bibnamefont
  {Sanquer}},\ }\href {http://dx.doi.org/10.1038/ncomms1140} {\bibfield
  {journal} {\bibinfo  {journal} {Nature Communications}\ }\textbf {\bibinfo
  {volume} {1}},\ \bibinfo {pages} {140 EP } (\bibinfo {year}
  {2010})}\BibitemShut {NoStop}%
\bibitem [{\citenamefont {Mondal}\ \emph
  {et~al.}(2011{\natexlab{a}})\citenamefont {Mondal}, \citenamefont
  {Kamlapure}, \citenamefont {Chand}, \citenamefont {Saraswat}, \citenamefont
  {Kumar}, \citenamefont {Jesudasan}, \citenamefont {Benfatto}, \citenamefont
  {Tripathi},\ and\ \citenamefont {Raychaudhuri}}]{MondalPRL11}%
  \BibitemOpen
  \bibfield  {author} {\bibinfo {author} {\bibfnamefont {M.}~\bibnamefont
  {Mondal}}, \bibinfo {author} {\bibfnamefont {A.}~\bibnamefont {Kamlapure}},
  \bibinfo {author} {\bibfnamefont {M.}~\bibnamefont {Chand}}, \bibinfo
  {author} {\bibfnamefont {G.}~\bibnamefont {Saraswat}}, \bibinfo {author}
  {\bibfnamefont {S.}~\bibnamefont {Kumar}}, \bibinfo {author} {\bibfnamefont
  {J.}~\bibnamefont {Jesudasan}}, \bibinfo {author} {\bibfnamefont
  {L.}~\bibnamefont {Benfatto}}, \bibinfo {author} {\bibfnamefont
  {V.}~\bibnamefont {Tripathi}}, \ and\ \bibinfo {author} {\bibfnamefont
  {P.}~\bibnamefont {Raychaudhuri}},\ }\href {\doibase
  10.1103/PhysRevLett.106.047001} {\bibfield  {journal} {\bibinfo  {journal}
  {Phys. Rev. Lett.}\ }\textbf {\bibinfo {volume} {106}},\ \bibinfo {pages}
  {047001} (\bibinfo {year} {2011}{\natexlab{a}})}\BibitemShut {NoStop}%
\bibitem [{\citenamefont {Chand}\ \emph {et~al.}(2012)\citenamefont {Chand},
  \citenamefont {Saraswat}, \citenamefont {Kamlapure}, \citenamefont {Mondal},
  \citenamefont {Kumar}, \citenamefont {Jesudasan}, \citenamefont {Bagwe},
  \citenamefont {Benfatto}, \citenamefont {Tripathi},\ and\ \citenamefont
  {Raychaudhuri}}]{ChandPRB2012}%
  \BibitemOpen
  \bibfield  {author} {\bibinfo {author} {\bibfnamefont {M.}~\bibnamefont
  {Chand}}, \bibinfo {author} {\bibfnamefont {G.}~\bibnamefont {Saraswat}},
  \bibinfo {author} {\bibfnamefont {A.}~\bibnamefont {Kamlapure}}, \bibinfo
  {author} {\bibfnamefont {M.}~\bibnamefont {Mondal}}, \bibinfo {author}
  {\bibfnamefont {S.}~\bibnamefont {Kumar}}, \bibinfo {author} {\bibfnamefont
  {J.}~\bibnamefont {Jesudasan}}, \bibinfo {author} {\bibfnamefont
  {V.}~\bibnamefont {Bagwe}}, \bibinfo {author} {\bibfnamefont
  {L.}~\bibnamefont {Benfatto}}, \bibinfo {author} {\bibfnamefont
  {V.}~\bibnamefont {Tripathi}}, \ and\ \bibinfo {author} {\bibfnamefont
  {P.}~\bibnamefont {Raychaudhuri}},\ }\href {\doibase
  10.1103/PhysRevB.85.014508} {\bibfield  {journal} {\bibinfo  {journal} {Phys.
  Rev. B}\ }\textbf {\bibinfo {volume} {85}},\ \bibinfo {pages} {014508}
  (\bibinfo {year} {2012})}\BibitemShut {NoStop}%
\bibitem [{\citenamefont {Breznay}\ \emph {et~al.}(2012)\citenamefont
  {Breznay}, \citenamefont {Michaeli}, \citenamefont {Tikhonov}, \citenamefont
  {Finkel'stein}, \citenamefont {Tendulkar},\ and\ \citenamefont
  {Kapitulnik}}]{BreznayPRB2012}%
  \BibitemOpen
  \bibfield  {author} {\bibinfo {author} {\bibfnamefont {N.~P.}\ \bibnamefont
  {Breznay}}, \bibinfo {author} {\bibfnamefont {K.}~\bibnamefont {Michaeli}},
  \bibinfo {author} {\bibfnamefont {K.~S.}\ \bibnamefont {Tikhonov}}, \bibinfo
  {author} {\bibfnamefont {A.~M.}\ \bibnamefont {Finkel'stein}}, \bibinfo
  {author} {\bibfnamefont {M.}~\bibnamefont {Tendulkar}}, \ and\ \bibinfo
  {author} {\bibfnamefont {A.}~\bibnamefont {Kapitulnik}},\ }\href {\doibase
  10.1103/PhysRevB.86.014514} {\bibfield  {journal} {\bibinfo  {journal} {Phys.
  Rev. B}\ }\textbf {\bibinfo {volume} {86}},\ \bibinfo {pages} {014514}
  (\bibinfo {year} {2012})}\BibitemShut {NoStop}%
\bibitem [{\citenamefont {Breznay}\ and\ \citenamefont
  {Kapitulnik}(2013)}]{BreznayPRB13}%
  \BibitemOpen
  \bibfield  {author} {\bibinfo {author} {\bibfnamefont {N.~P.}\ \bibnamefont
  {Breznay}}\ and\ \bibinfo {author} {\bibfnamefont {A.}~\bibnamefont
  {Kapitulnik}},\ }\href {\doibase 10.1103/PhysRevB.88.104510} {\bibfield
  {journal} {\bibinfo  {journal} {Phys. Rev. B}\ }\textbf {\bibinfo {volume}
  {88}},\ \bibinfo {pages} {104510} (\bibinfo {year} {2013})}\BibitemShut
  {NoStop}%
\bibitem [{\citenamefont {Michaeli}\ \emph {et~al.}(2012)\citenamefont
  {Michaeli}, \citenamefont {Tikhonov},\ and\ \citenamefont
  {Finkel'stein}}]{MichaeliPRB2012}%
  \BibitemOpen
  \bibfield  {author} {\bibinfo {author} {\bibfnamefont {K.}~\bibnamefont
  {Michaeli}}, \bibinfo {author} {\bibfnamefont {K.~S.}\ \bibnamefont
  {Tikhonov}}, \ and\ \bibinfo {author} {\bibfnamefont {A.~M.}\ \bibnamefont
  {Finkel'stein}},\ }\href {\doibase 10.1103/PhysRevB.86.014515} {\bibfield
  {journal} {\bibinfo  {journal} {Phys. Rev. B}\ }\textbf {\bibinfo {volume}
  {86}},\ \bibinfo {pages} {014515} (\bibinfo {year} {2012})}\BibitemShut
  {NoStop}%
\bibitem [{\citenamefont {Chockalingam}\ \emph {et~al.}(2008)\citenamefont
  {Chockalingam}, \citenamefont {Chand}, \citenamefont {Jesudasan},
  \citenamefont {Tripathi},\ and\ \citenamefont
  {Raychaudhuri}}]{ChockalingamPRB2008}%
  \BibitemOpen
  \bibfield  {author} {\bibinfo {author} {\bibfnamefont {S.~P.}\ \bibnamefont
  {Chockalingam}}, \bibinfo {author} {\bibfnamefont {M.}~\bibnamefont {Chand}},
  \bibinfo {author} {\bibfnamefont {J.}~\bibnamefont {Jesudasan}}, \bibinfo
  {author} {\bibfnamefont {V.}~\bibnamefont {Tripathi}}, \ and\ \bibinfo
  {author} {\bibfnamefont {P.}~\bibnamefont {Raychaudhuri}},\ }\href {\doibase
  10.1103/PhysRevB.77.214503} {\bibfield  {journal} {\bibinfo  {journal} {Phys.
  Rev. B}\ }\textbf {\bibinfo {volume} {77}},\ \bibinfo {pages} {214503}
  (\bibinfo {year} {2008})}\BibitemShut {NoStop}%
\bibitem [{\citenamefont {Chand}\ \emph {et~al.}(2009)\citenamefont {Chand},
  \citenamefont {Mishra}, \citenamefont {Xiong}, \citenamefont {Kamlapure},
  \citenamefont {Chockalingam}, \citenamefont {Jesudasan}, \citenamefont
  {Bagwe}, \citenamefont {Mondal}, \citenamefont {Adams}, \citenamefont
  {Tripathi},\ and\ \citenamefont {Raychaudhuri}}]{ChandPRB09}%
  \BibitemOpen
  \bibfield  {author} {\bibinfo {author} {\bibfnamefont {M.}~\bibnamefont
  {Chand}}, \bibinfo {author} {\bibfnamefont {A.}~\bibnamefont {Mishra}},
  \bibinfo {author} {\bibfnamefont {Y.~M.}\ \bibnamefont {Xiong}}, \bibinfo
  {author} {\bibfnamefont {A.}~\bibnamefont {Kamlapure}}, \bibinfo {author}
  {\bibfnamefont {S.~P.}\ \bibnamefont {Chockalingam}}, \bibinfo {author}
  {\bibfnamefont {J.}~\bibnamefont {Jesudasan}}, \bibinfo {author}
  {\bibfnamefont {V.}~\bibnamefont {Bagwe}}, \bibinfo {author} {\bibfnamefont
  {M.}~\bibnamefont {Mondal}}, \bibinfo {author} {\bibfnamefont {P.~W.}\
  \bibnamefont {Adams}}, \bibinfo {author} {\bibfnamefont {V.}~\bibnamefont
  {Tripathi}}, \ and\ \bibinfo {author} {\bibfnamefont {P.}~\bibnamefont
  {Raychaudhuri}},\ }\href {\doibase 10.1103/PhysRevB.80.134514} {\bibfield
  {journal} {\bibinfo  {journal} {Phys. Rev. B}\ }\textbf {\bibinfo {volume}
  {80}},\ \bibinfo {pages} {134514} (\bibinfo {year} {2009})}\BibitemShut
  {NoStop}%
\bibitem [{\citenamefont {Semenov}\ \emph {et~al.}(2009)\citenamefont
  {Semenov}, \citenamefont {G\"unther}, \citenamefont {B\"ottger},
  \citenamefont {H\"ubers}, \citenamefont {Bartolf}, \citenamefont {Engel},
  \citenamefont {Schilling}, \citenamefont {Ilin}, \citenamefont {Siegel},
  \citenamefont {Schneider}, \citenamefont {Gerthsen},\ and\ \citenamefont
  {Gippius}}]{SemenovPRB2009}%
  \BibitemOpen
  \bibfield  {author} {\bibinfo {author} {\bibfnamefont {A.}~\bibnamefont
  {Semenov}}, \bibinfo {author} {\bibfnamefont {B.}~\bibnamefont {G\"unther}},
  \bibinfo {author} {\bibfnamefont {U.}~\bibnamefont {B\"ottger}}, \bibinfo
  {author} {\bibfnamefont {H.-W.}\ \bibnamefont {H\"ubers}}, \bibinfo {author}
  {\bibfnamefont {H.}~\bibnamefont {Bartolf}}, \bibinfo {author} {\bibfnamefont
  {A.}~\bibnamefont {Engel}}, \bibinfo {author} {\bibfnamefont
  {A.}~\bibnamefont {Schilling}}, \bibinfo {author} {\bibfnamefont
  {K.}~\bibnamefont {Ilin}}, \bibinfo {author} {\bibfnamefont {M.}~\bibnamefont
  {Siegel}}, \bibinfo {author} {\bibfnamefont {R.}~\bibnamefont {Schneider}},
  \bibinfo {author} {\bibfnamefont {D.}~\bibnamefont {Gerthsen}}, \ and\
  \bibinfo {author} {\bibfnamefont {N.~A.}\ \bibnamefont {Gippius}},\ }\href
  {\doibase 10.1103/PhysRevB.80.054510} {\bibfield  {journal} {\bibinfo
  {journal} {Phys. Rev. B}\ }\textbf {\bibinfo {volume} {80}},\ \bibinfo
  {pages} {054510} (\bibinfo {year} {2009})}\BibitemShut {NoStop}%
\bibitem [{\citenamefont {Beck}\ \emph {et~al.}(2011)\citenamefont {Beck},
  \citenamefont {Klammer}, \citenamefont {Lang}, \citenamefont {Leiderer},
  \citenamefont {Kabanov}, \citenamefont {Gol'tsman},\ and\ \citenamefont
  {Demsar}}]{BeckPRL11}%
  \BibitemOpen
  \bibfield  {author} {\bibinfo {author} {\bibfnamefont {M.}~\bibnamefont
  {Beck}}, \bibinfo {author} {\bibfnamefont {M.}~\bibnamefont {Klammer}},
  \bibinfo {author} {\bibfnamefont {S.}~\bibnamefont {Lang}}, \bibinfo {author}
  {\bibfnamefont {P.}~\bibnamefont {Leiderer}}, \bibinfo {author}
  {\bibfnamefont {V.~V.}\ \bibnamefont {Kabanov}}, \bibinfo {author}
  {\bibfnamefont {G.~N.}\ \bibnamefont {Gol'tsman}}, \ and\ \bibinfo {author}
  {\bibfnamefont {J.}~\bibnamefont {Demsar}},\ }\href {\doibase
  10.1103/PhysRevLett.107.177007} {\bibfield  {journal} {\bibinfo  {journal}
  {Phys. Rev. Lett.}\ }\textbf {\bibinfo {volume} {107}},\ \bibinfo {pages}
  {177007} (\bibinfo {year} {2011})}\BibitemShut {NoStop}%
\bibitem [{\citenamefont {Matsunaga}\ and\ \citenamefont
  {Shimano}(2012)}]{MatsunagaPRL12}%
  \BibitemOpen
  \bibfield  {author} {\bibinfo {author} {\bibfnamefont {R.}~\bibnamefont
  {Matsunaga}}\ and\ \bibinfo {author} {\bibfnamefont {R.}~\bibnamefont
  {Shimano}},\ }\href {\doibase 10.1103/PhysRevLett.109.187002} {\bibfield
  {journal} {\bibinfo  {journal} {Phys. Rev. Lett.}\ }\textbf {\bibinfo
  {volume} {109}},\ \bibinfo {pages} {187002} (\bibinfo {year}
  {2012})}\BibitemShut {NoStop}%
\bibitem [{\citenamefont {Beck}\ \emph {et~al.}(2013)\citenamefont {Beck},
  \citenamefont {Rousseau}, \citenamefont {Klammer}, \citenamefont {Leiderer},
  \citenamefont {Mittendorff}, \citenamefont {Winnerl}, \citenamefont {Helm},
  \citenamefont {Gol'tsman},\ and\ \citenamefont {Demsar}}]{BeckPRL13}%
  \BibitemOpen
  \bibfield  {author} {\bibinfo {author} {\bibfnamefont {M.}~\bibnamefont
  {Beck}}, \bibinfo {author} {\bibfnamefont {I.}~\bibnamefont {Rousseau}},
  \bibinfo {author} {\bibfnamefont {M.}~\bibnamefont {Klammer}}, \bibinfo
  {author} {\bibfnamefont {P.}~\bibnamefont {Leiderer}}, \bibinfo {author}
  {\bibfnamefont {M.}~\bibnamefont {Mittendorff}}, \bibinfo {author}
  {\bibfnamefont {S.}~\bibnamefont {Winnerl}}, \bibinfo {author} {\bibfnamefont
  {M.}~\bibnamefont {Helm}}, \bibinfo {author} {\bibfnamefont {G.~N.}\
  \bibnamefont {Gol'tsman}}, \ and\ \bibinfo {author} {\bibfnamefont
  {J.}~\bibnamefont {Demsar}},\ }\href {\doibase
  10.1103/PhysRevLett.110.267003} {\bibfield  {journal} {\bibinfo  {journal}
  {Phys. Rev. Lett.}\ }\textbf {\bibinfo {volume} {110}},\ \bibinfo {pages}
  {267003} (\bibinfo {year} {2013})}\BibitemShut {NoStop}%
\bibitem [{\citenamefont {Matsunaga}\ \emph {et~al.}(2014)\citenamefont
  {Matsunaga}, \citenamefont {Tsuji}, \citenamefont {Fujita}, \citenamefont
  {Sugioka}, \citenamefont {Makise}, \citenamefont {Uzawa}, \citenamefont
  {Terai}, \citenamefont {Wang}, \citenamefont {Aoki},\ and\ \citenamefont
  {Shimano}}]{MatsunagaSCIENCE14}%
  \BibitemOpen
  \bibfield  {author} {\bibinfo {author} {\bibfnamefont {R.}~\bibnamefont
  {Matsunaga}}, \bibinfo {author} {\bibfnamefont {N.}~\bibnamefont {Tsuji}},
  \bibinfo {author} {\bibfnamefont {H.}~\bibnamefont {Fujita}}, \bibinfo
  {author} {\bibfnamefont {A.}~\bibnamefont {Sugioka}}, \bibinfo {author}
  {\bibfnamefont {K.}~\bibnamefont {Makise}}, \bibinfo {author} {\bibfnamefont
  {Y.}~\bibnamefont {Uzawa}}, \bibinfo {author} {\bibfnamefont
  {H.}~\bibnamefont {Terai}}, \bibinfo {author} {\bibfnamefont
  {Z.}~\bibnamefont {Wang}}, \bibinfo {author} {\bibfnamefont {H.}~\bibnamefont
  {Aoki}}, \ and\ \bibinfo {author} {\bibfnamefont {R.}~\bibnamefont
  {Shimano}},\ }\href {\doibase 10.1126/science.1254697} {\bibfield  {journal}
  {\bibinfo  {journal} {Science}\ }\textbf {\bibinfo {volume} {345}},\ \bibinfo
  {pages} {1145} (\bibinfo {year} {2014})}\BibitemShut {NoStop}%
\bibitem [{\citenamefont {Aronov}\ \emph {et~al.}(1995)\citenamefont {Aronov},
  \citenamefont {Hikami},\ and\ \citenamefont {Larkin}}]{AronovPRB95}%
  \BibitemOpen
  \bibfield  {author} {\bibinfo {author} {\bibfnamefont {A.~G.}\ \bibnamefont
  {Aronov}}, \bibinfo {author} {\bibfnamefont {S.}~\bibnamefont {Hikami}}, \
  and\ \bibinfo {author} {\bibfnamefont {A.~I.}\ \bibnamefont {Larkin}},\
  }\href {\doibase 10.1103/PhysRevB.51.3880} {\bibfield  {journal} {\bibinfo
  {journal} {Phys. Rev. B}\ }\textbf {\bibinfo {volume} {51}},\ \bibinfo
  {pages} {3880} (\bibinfo {year} {1995})}\BibitemShut {NoStop}%
\bibitem [{\citenamefont {Werthamer}\ \emph {et~al.}(1966)\citenamefont
  {Werthamer}, \citenamefont {Helfand},\ and\ \citenamefont
  {Hohenberg}}]{WerthamerPRB1966}%
  \BibitemOpen
  \bibfield  {author} {\bibinfo {author} {\bibfnamefont {N.~R.}\ \bibnamefont
  {Werthamer}}, \bibinfo {author} {\bibfnamefont {E.}~\bibnamefont {Helfand}},
  \ and\ \bibinfo {author} {\bibfnamefont {P.~C.}\ \bibnamefont {Hohenberg}},\
  }\href {\doibase 10.1103/PhysRev.147.295} {\bibfield  {journal} {\bibinfo
  {journal} {Phys. Rev.}\ }\textbf {\bibinfo {volume} {147}},\ \bibinfo {pages}
  {295} (\bibinfo {year} {1966})}\BibitemShut {NoStop}%
\bibitem [{\citenamefont {Mondal}\ \emph
  {et~al.}(2011{\natexlab{b}})\citenamefont {Mondal}, \citenamefont {Kumar},
  \citenamefont {Chand}, \citenamefont {Kamlapure}, \citenamefont {Saraswat},
  \citenamefont {Seibold}, \citenamefont {Benfatto},\ and\ \citenamefont
  {Raychaudhuri}}]{MondalPRL11a}%
  \BibitemOpen
  \bibfield  {author} {\bibinfo {author} {\bibfnamefont {M.}~\bibnamefont
  {Mondal}}, \bibinfo {author} {\bibfnamefont {S.}~\bibnamefont {Kumar}},
  \bibinfo {author} {\bibfnamefont {M.}~\bibnamefont {Chand}}, \bibinfo
  {author} {\bibfnamefont {A.}~\bibnamefont {Kamlapure}}, \bibinfo {author}
  {\bibfnamefont {G.}~\bibnamefont {Saraswat}}, \bibinfo {author}
  {\bibfnamefont {G.}~\bibnamefont {Seibold}}, \bibinfo {author} {\bibfnamefont
  {L.}~\bibnamefont {Benfatto}}, \ and\ \bibinfo {author} {\bibfnamefont
  {P.}~\bibnamefont {Raychaudhuri}},\ }\href {\doibase
  10.1103/PhysRevLett.107.217003} {\bibfield  {journal} {\bibinfo  {journal}
  {Phys. Rev. Lett.}\ }\textbf {\bibinfo {volume} {107}},\ \bibinfo {pages}
  {217003} (\bibinfo {year} {2011}{\natexlab{b}})}\BibitemShut {NoStop}%
\bibitem [{\citenamefont {Halperin}\ and\ \citenamefont
  {Nelson}(1979)}]{HalperinJLTP1979}%
  \BibitemOpen
  \bibfield  {author} {\bibinfo {author} {\bibfnamefont {B.~I.}\ \bibnamefont
  {Halperin}}\ and\ \bibinfo {author} {\bibfnamefont {D.~R.}\ \bibnamefont
  {Nelson}},\ }\href {\doibase 10.1007/BF00116988} {\bibfield  {journal}
  {\bibinfo  {journal} {Journal of Low Temperature Physics}\ }\textbf {\bibinfo
  {volume} {36}},\ \bibinfo {pages} {599} (\bibinfo {year} {1979})}\BibitemShut
  {NoStop}%
\bibitem [{\citenamefont {Benfatto}\ \emph {et~al.}(2009)\citenamefont
  {Benfatto}, \citenamefont {Castellani},\ and\ \citenamefont
  {Giamarchi}}]{BenfattoPRB2009}%
  \BibitemOpen
  \bibfield  {author} {\bibinfo {author} {\bibfnamefont {L.}~\bibnamefont
  {Benfatto}}, \bibinfo {author} {\bibfnamefont {C.}~\bibnamefont
  {Castellani}}, \ and\ \bibinfo {author} {\bibfnamefont {T.}~\bibnamefont
  {Giamarchi}},\ }\href {\doibase 10.1103/PhysRevB.80.214506} {\bibfield
  {journal} {\bibinfo  {journal} {Phys. Rev. B}\ }\textbf {\bibinfo {volume}
  {80}},\ \bibinfo {pages} {214506} (\bibinfo {year} {2009})}\BibitemShut
  {NoStop}%
\bibitem [{\citenamefont {Pourret}\ \emph {et~al.}(2007)\citenamefont
  {Pourret}, \citenamefont {Aubin}, \citenamefont {Lesueur}, \citenamefont
  {Marrache-Kikuchi}, \citenamefont {Berg\'e}, \citenamefont {Dumoulin},\ and\
  \citenamefont {Behnia}}]{PourretPRB07}%
  \BibitemOpen
  \bibfield  {author} {\bibinfo {author} {\bibfnamefont {A.}~\bibnamefont
  {Pourret}}, \bibinfo {author} {\bibfnamefont {H.}~\bibnamefont {Aubin}},
  \bibinfo {author} {\bibfnamefont {J.}~\bibnamefont {Lesueur}}, \bibinfo
  {author} {\bibfnamefont {C.~A.}\ \bibnamefont {Marrache-Kikuchi}}, \bibinfo
  {author} {\bibfnamefont {L.}~\bibnamefont {Berg\'e}}, \bibinfo {author}
  {\bibfnamefont {L.}~\bibnamefont {Dumoulin}}, \ and\ \bibinfo {author}
  {\bibfnamefont {K.}~\bibnamefont {Behnia}},\ }\href {\doibase
  10.1103/PhysRevB.76.214504} {\bibfield  {journal} {\bibinfo  {journal} {Phys.
  Rev. B}\ }\textbf {\bibinfo {volume} {76}},\ \bibinfo {pages} {214504}
  (\bibinfo {year} {2007})}\BibitemShut {NoStop}%
\bibitem [{\citenamefont {Pourret}\ \emph {et~al.}(2009)\citenamefont
  {Pourret}, \citenamefont {Spathis}, \citenamefont {Aubin},\ and\
  \citenamefont {Behnia}}]{PourretNJP09}%
  \BibitemOpen
  \bibfield  {author} {\bibinfo {author} {\bibfnamefont {A.}~\bibnamefont
  {Pourret}}, \bibinfo {author} {\bibfnamefont {P.}~\bibnamefont {Spathis}},
  \bibinfo {author} {\bibfnamefont {H.}~\bibnamefont {Aubin}}, \ and\ \bibinfo
  {author} {\bibfnamefont {K.}~\bibnamefont {Behnia}},\ }\href
  {http://stacks.iop.org/1367-2630/11/i=5/a=055071} {\bibfield  {journal}
  {\bibinfo  {journal} {New Journal of Physics}\ }\textbf {\bibinfo {volume}
  {11}},\ \bibinfo {pages} {055071} (\bibinfo {year} {2009})}\BibitemShut
  {NoStop}%
\bibitem [{\citenamefont {Tafti}\ \emph {et~al.}(2014)\citenamefont {Tafti},
  \citenamefont {Lalibert\'e}, \citenamefont {Dion}, \citenamefont {Gaudet},
  \citenamefont {Fournier},\ and\ \citenamefont {Taillefer}}]{TaftiPRB14}%
  \BibitemOpen
  \bibfield  {author} {\bibinfo {author} {\bibfnamefont {F.~F.}\ \bibnamefont
  {Tafti}}, \bibinfo {author} {\bibfnamefont {F.}~\bibnamefont {Lalibert\'e}},
  \bibinfo {author} {\bibfnamefont {M.}~\bibnamefont {Dion}}, \bibinfo {author}
  {\bibfnamefont {J.}~\bibnamefont {Gaudet}}, \bibinfo {author} {\bibfnamefont
  {P.}~\bibnamefont {Fournier}}, \ and\ \bibinfo {author} {\bibfnamefont
  {L.}~\bibnamefont {Taillefer}},\ }\href {\doibase 10.1103/PhysRevB.90.024519}
  {\bibfield  {journal} {\bibinfo  {journal} {Phys. Rev. B}\ }\textbf {\bibinfo
  {volume} {90}},\ \bibinfo {pages} {024519} (\bibinfo {year}
  {2014})}\BibitemShut {NoStop}%
\bibitem [{\citenamefont {Yamashita}\ \emph {et~al.}(2015)\citenamefont
  {Yamashita}, \citenamefont {Shimoyama}, \citenamefont {Haga}, \citenamefont
  {Matsuda}, \citenamefont {Yamamoto}, \citenamefont {Onuki}, \citenamefont
  {Sumiyoshi}, \citenamefont {Fujimoto}, \citenamefont {Levchenko},
  \citenamefont {Shibauchi},\ and\ \citenamefont {Matsuda}}]{Yamashita2015}%
  \BibitemOpen
  \bibfield  {author} {\bibinfo {author} {\bibfnamefont {T.}~\bibnamefont
  {Yamashita}}, \bibinfo {author} {\bibfnamefont {Y.}~\bibnamefont
  {Shimoyama}}, \bibinfo {author} {\bibfnamefont {Y.}~\bibnamefont {Haga}},
  \bibinfo {author} {\bibfnamefont {T.~D.}\ \bibnamefont {Matsuda}}, \bibinfo
  {author} {\bibfnamefont {E.}~\bibnamefont {Yamamoto}}, \bibinfo {author}
  {\bibfnamefont {Y.}~\bibnamefont {Onuki}}, \bibinfo {author} {\bibfnamefont
  {H.}~\bibnamefont {Sumiyoshi}}, \bibinfo {author} {\bibfnamefont
  {S.}~\bibnamefont {Fujimoto}}, \bibinfo {author} {\bibfnamefont
  {A.}~\bibnamefont {Levchenko}}, \bibinfo {author} {\bibfnamefont
  {T.}~\bibnamefont {Shibauchi}}, \ and\ \bibinfo {author} {\bibfnamefont
  {Y.}~\bibnamefont {Matsuda}},\ }\href {http://dx.doi.org/10.1038/nphys3170}
  {\bibfield  {journal} {\bibinfo  {journal} {Nat Phys}\ }\textbf {\bibinfo
  {volume} {11}},\ \bibinfo {pages} {17} (\bibinfo {year} {2015})}\BibitemShut
  {NoStop}%
\bibitem [{\citenamefont {Tikhonov}\ \emph {et~al.}(2012)\citenamefont
  {Tikhonov}, \citenamefont {Schwiete},\ and\ \citenamefont
  {Finkel'stein}}]{TikhonovPRB12}%
  \BibitemOpen
  \bibfield  {author} {\bibinfo {author} {\bibfnamefont {K.~S.}\ \bibnamefont
  {Tikhonov}}, \bibinfo {author} {\bibfnamefont {G.}~\bibnamefont {Schwiete}},
  \ and\ \bibinfo {author} {\bibfnamefont {A.~M.}\ \bibnamefont
  {Finkel'stein}},\ }\href {\doibase 10.1103/PhysRevB.85.174527} {\bibfield
  {journal} {\bibinfo  {journal} {Phys. Rev. B}\ }\textbf {\bibinfo {volume}
  {85}},\ \bibinfo {pages} {174527} (\bibinfo {year} {2012})}\BibitemShut
  {NoStop}%
\bibitem [{\citenamefont {Aslamazov}\ and\ \citenamefont
  {Larkin}(1968)}]{Aslamazov1968}%
  \BibitemOpen
  \bibfield  {author} {\bibinfo {author} {\bibfnamefont {L.~G.}\ \bibnamefont
  {Aslamazov}}\ and\ \bibinfo {author} {\bibfnamefont {A.~I.}\ \bibnamefont
  {Larkin}},\ }\href@noop {} {\bibfield  {journal} {\bibinfo  {journal} {Sov.
  Phys. Solid State}\ }\textbf {\bibinfo {volume} {10}},\ \bibinfo {pages}
  {875} (\bibinfo {year} {1968})}\BibitemShut {NoStop}%
\bibitem [{\citenamefont {Larkin}\ and\ \citenamefont
  {Varlamov}(2005)}]{VarlamovBook}%
  \BibitemOpen
  \bibfield  {author} {\bibinfo {author} {\bibfnamefont {A.}~\bibnamefont
  {Larkin}}\ and\ \bibinfo {author} {\bibfnamefont {A.}~\bibnamefont
  {Varlamov}},\ }\href@noop {} {\emph {\bibinfo {title} {Theory of Fluctuations
  in Superconductors}}}\ (\bibinfo  {publisher} {Oxford University Press},\
  \bibinfo {year} {2005})\BibitemShut {NoStop}%
\bibitem [{\citenamefont {Maki}(1968)}]{Maki68}%
  \BibitemOpen
  \bibfield  {author} {\bibinfo {author} {\bibfnamefont {K.}~\bibnamefont
  {Maki}},\ }\href {\doibase 10.1143/PTP.39.897} {\bibfield  {journal}
  {\bibinfo  {journal} {Progress of Theoretical Physics}\ }\textbf {\bibinfo
  {volume} {39}},\ \bibinfo {pages} {897} (\bibinfo {year} {1968})}\BibitemShut
  {NoStop}%
\bibitem [{\citenamefont {Thompson}(1970)}]{ThompsonPRB70}%
  \BibitemOpen
  \bibfield  {author} {\bibinfo {author} {\bibfnamefont {R.~S.}\ \bibnamefont
  {Thompson}},\ }\href {\doibase 10.1103/PhysRevB.1.327} {\bibfield  {journal}
  {\bibinfo  {journal} {Phys. Rev. B}\ }\textbf {\bibinfo {volume} {1}},\
  \bibinfo {pages} {327} (\bibinfo {year} {1970})}\BibitemShut {NoStop}%
\bibitem [{\citenamefont {Lang}\ \emph {et~al.}(1994)\citenamefont {Lang},
  \citenamefont {Heine}, \citenamefont {Schwab}, \citenamefont {Wang},\ and\
  \citenamefont {B\"auerle}}]{LangPRB94}%
  \BibitemOpen
  \bibfield  {author} {\bibinfo {author} {\bibfnamefont {W.}~\bibnamefont
  {Lang}}, \bibinfo {author} {\bibfnamefont {G.}~\bibnamefont {Heine}},
  \bibinfo {author} {\bibfnamefont {P.}~\bibnamefont {Schwab}}, \bibinfo
  {author} {\bibfnamefont {X.~Z.}\ \bibnamefont {Wang}}, \ and\ \bibinfo
  {author} {\bibfnamefont {D.}~\bibnamefont {B\"auerle}},\ }\href {\doibase
  10.1103/PhysRevB.49.4209} {\bibfield  {journal} {\bibinfo  {journal} {Phys.
  Rev. B}\ }\textbf {\bibinfo {volume} {49}},\ \bibinfo {pages} {4209}
  (\bibinfo {year} {1994})}\BibitemShut {NoStop}%
\bibitem [{\citenamefont {Reizer}(1992)}]{ReizerPRB92}%
  \BibitemOpen
  \bibfield  {author} {\bibinfo {author} {\bibfnamefont {M.~Y.}\ \bibnamefont
  {Reizer}},\ }\href {\doibase 10.1103/PhysRevB.45.12949} {\bibfield  {journal}
  {\bibinfo  {journal} {Phys. Rev. B}\ }\textbf {\bibinfo {volume} {45}},\
  \bibinfo {pages} {12949} (\bibinfo {year} {1992})}\BibitemShut {NoStop}%
\bibitem [{\citenamefont {Bardeen}\ \emph {et~al.}(1957)\citenamefont
  {Bardeen}, \citenamefont {Cooper},\ and\ \citenamefont
  {Schrieffer}}]{BardeenPR1957}%
  \BibitemOpen
  \bibfield  {author} {\bibinfo {author} {\bibfnamefont {J.}~\bibnamefont
  {Bardeen}}, \bibinfo {author} {\bibfnamefont {L.~N.}\ \bibnamefont {Cooper}},
  \ and\ \bibinfo {author} {\bibfnamefont {J.~R.}\ \bibnamefont {Schrieffer}},\
  }\href {\doibase 10.1103/PhysRev.108.1175} {\bibfield  {journal} {\bibinfo
  {journal} {Phys. Rev.}\ }\textbf {\bibinfo {volume} {108}},\ \bibinfo {pages}
  {1175} (\bibinfo {year} {1957})}\BibitemShut {NoStop}%
\bibitem [{\citenamefont {McMillan}(1968)}]{McMillanPR68}%
  \BibitemOpen
  \bibfield  {author} {\bibinfo {author} {\bibfnamefont {W.~L.}\ \bibnamefont
  {McMillan}},\ }\href {\doibase 10.1103/PhysRev.167.331} {\bibfield  {journal}
  {\bibinfo  {journal} {Phys. Rev.}\ }\textbf {\bibinfo {volume} {167}},\
  \bibinfo {pages} {331} (\bibinfo {year} {1968})}\BibitemShut {NoStop}%
\end{thebibliography}
%

\end{document}